\documentclass[aps,pra,reprint,superscriptaddress,longbibliography,noeprint]{revtex4-1} 
\usepackage[colorlinks,pdfstartview=FitH,citecolor=blue,linkcolor=blue,urlcolor=blue]{hyperref} 
\usepackage{graphicx}
\usepackage{amsmath}
\usepackage{amssymb}
\usepackage{bm}
\usepackage{mhchem} 
\usepackage{graphicx}
\usepackage{siunitx}
\usepackage{bm} 

\renewcommand{\Re}{\operatorname{Re}}
\renewcommand{\Im}{\operatorname{Im}}
\newcommand{\citeasnoun}[1]{Ref.~\cite{#1}} 

\renewcommand{\eqref}[1]{Eq.~(\ref{eq:#1})}
\newcommand{\eqreftwo}[2]{Eqs.~(\ref{eq:#1},\ref{eq:#2})}
\newcommand{\eqrefrange}[2]{Eqs.~(\ref{eq:#1}--\ref{eq:#2})}
\newcommand{\Eqref}[1]{Equation~(\ref{eq:#1})}
\newcommand{\figref}[1]{Fig.~\ref{fig:#1}}
\newcommand{\Figref}[1]{Figure~\ref{fig:#1}}
\newcommand{\cc}[1]{#1^*}
\newcommand{\vect}[1]{\mathbf{#1}}
\newcommand*{\tens}[1]{\bm{#1}}
\newcommand*{\Pabs}{P_{\rm abs}}
\newcommand*{\Pscat}{P_{\rm scat}}
\newcommand*{\Pext}{P_{\rm ext}}
\newcommand*{\Ev}{\vect{E}}
\newcommand*{\Kv}{\vect{K}}
\newcommand*{\xv}{\vect{x}}
\newcommand*{\sigmat}{\tens{\sigma}}
\newcommand*{\EF}{E_{\rm F}}
\newcommand*{\omegap}{\omega_{\rm p}}
\newcommand*{\SM}{Supp. Info.}
\DeclareMathOperator{\Lke}{\mathcal{L}}

\begin{document} 

\title{Limits to the Optical Response of Graphene and 2D Materials}

\author{Owen D. Miller}
\email[Corresponding author: ]{owen.miller@yale.edu}  
\affiliation{Department of Applied Physics and Energy Sciences Institute, Yale University, New Haven, CT 06511}
\author{Ognjen Ilic}
\affiliation{Department of Applied Physics and Material Science, California Institute of Technology, Pasadena, CA 91125}
\author{Thomas Christensen}
\affiliation{Department of Physics, Massachusetts Institute of Technology, Cambridge, MA 02139}
\author{M. T. Homer Reid}
\affiliation{Department of Mathematics, Massachusetts Institute of Technology, Cambridge, MA 02139}
\author{Harry~A.~Atwater}
\affiliation{Department of Applied Physics and Material Science, California Institute of Technology, Pasadena, CA 91125}
\author{John D. Joannopoulos}
\affiliation{Department of Physics, Massachusetts Institute of Technology, Cambridge, MA 02139}
\author{Marin Solja\v{c}i\'{c}}
\affiliation{Department of Physics, Massachusetts Institute of Technology, Cambridge, MA 02139}
\author{Steven G. Johnson}
\affiliation{Department of Physics, Massachusetts Institute of Technology, Cambridge, MA 02139}
\affiliation{Department of Mathematics, Massachusetts Institute of Technology, Cambridge, MA 02139}


\begin{abstract}
    2D materials provide a platform for strong light--matter interactions, creating wide-ranging design opportunities via new-material discoveries and new methods for geometrical structuring. We derive general upper bounds to the strength of such light--matter interactions, given only the optical conductivity of the material, including spatial nonlocality, and otherwise independent of shape and configuration. Our material figure of merit shows that highly doped graphene is an optimal material at infrared frequencies, whereas single-atomic-layer silver is optimal in the visible. For quantities ranging from absorption and scattering to near-field spontaneous-emission enhancements and radiative heat transfer, we consider canonical geometrical structures and show that in certain cases the bounds can be approached, while in others there may be significant opportunity for design improvement. The bounds can encourage systematic improvements in the design of ultrathin broadband absorbers, 2D antennas, and near-field energy harvesters. 
\end{abstract}

\maketitle
\sisetup{range-phrase=--}
\sisetup{range-units=single}
2D materials~\cite{Novoselov2005a,Geim2007} and emerging methods~\cite{Pang2009,He2010,Thongrattanasiri2012,Zhan2012,Piper2014,Cai2015} for patterning 2D layers and their surroundings are opening an expansive design space, exhibiting significantly different optical~\cite{Koppens2011,Basov2016,Low2016} (and electronic) properties from their 3D counterparts. In this Letter, we identify energy constraints embedded within Maxwell's equations that impose theoretical bounds on the largest optical response that can be generated in any 2D material, in the near or far field. The bounds account for material loss as encoded in the real part of a material's conductivity---in the case of a spatially local conductivity tensor $\sigmat$, they are proportional to $\big\|\sigmat^\dagger \left(\Re \sigmat\right)^{-1} \sigmat\big\|$---and are otherwise independent of shape and configuration. We derive the bounds through convex constraints imposed by the optical theorem~\cite{Newton1976,Jackson1999,Lytle2005} and its near-field analogue, leveraging a recent approach we developed for spatially local 3D materials~\cite{Miller2016}. In addition to accommodating nonlocal models, this work demonstrates starkly different near-field dependencies of 2D and 3D materials. For graphene, the 2D material of foremost interest to date, the bounds bifurcate into distinctive low- and high-energy regimes: the low-energy bounds are proportional to the Fermi level, whereas the high-energy bounds are proportional to the fine-structure constant, $\alpha$, for any geometrical configuration. We find that far-field bounds on the extinction cross-section can be approached by elliptical graphene disks, whereas the near-field bounds on the local density of states~\cite{Novotny2012,Joulain2003,Martin1998,DAguanno2004,OskooiJo13-sources} and radiative heat transfer rate~\cite{Polder1971,Rytov1988,Pendry1999,Mulet2002,Joulain2005,Volokitin2007} cannot be approached in prototypical flat-sheet configurations. The bounds presented here provide a simple material figure of merit to evaluate the emerging zoo of 2D materials, and offer the prospect of greater optical response via computational design. The material figure of merit can guide ongoing efforts in 2D-material discovery, while the general bounds can shape and drive efforts towards new levels of performance and better optical components.

Plasmonics in 2D materials opens the possibility for stronger light--matter interactions, which may be useful for technological applications, including single-molecule imaging~\cite{Nie1997,Kneipp1997,VanZanten2009,Schermelleh2010} and photovoltaics~\cite{Atwater2010,Ilic2012a}, as well as for basic-science discoveries, such as revealing forbidden transitions~\cite{Rivera2016}, and achieving unity optical absorption in graphene through optical impedance matching~\cite{Thongrattanasiri2012,Jang2014,Zhu2016,Kim2017}. Theoretical work towards understanding optical response in 2D materials has focused on analytical expressions using specific geometrical~\cite{Koppens2011,Thongrattanasiri2012,DeAbajo2014}  or metamaterial-based~\cite{Tassin2012} models, but from a design perspective such assumptions are restrictive. Quasistatic sum rules can yield upper limits on the cross-section~\cite{DeAbajo2015,Miller2014}, but have been restricted to far-field quantities and isotropic and spatially local materials. A well-known microwave-engineering bound, known as Rozanov's theorem~\cite{Rozanov2000}, offers a bandwidth limit as a function of material thickness, but its contour-integral approach requires perfectly conducting boundaries that are not applicable for 2D materials at optical frequencies. Here, we find constraints that do yield 2D-material optical-response bounds given only the material properties. We provide a general framework to derive limits to any optical-response quantity (including cross-sections, spontaneous-emission enhancements, and radiative-heat exchange),  and we present computational results suggesting pathways to approach the new bounds. For a broad class of hydrodynamic nonlocal-conductivity models~\cite{Ciraci2012,Mortensen2014}, which capture several important nonclassical features at length scales approaching the quantum regime, we derive general bounds in terms of a constitutive-relation operator. We show that the nonlocal response is necessarily bounded above by the local-response bounds; further, by exploiting the quasistatic nature of interactions at nonlocal length scales, we show that the maximum response must be reduced in proportion to a ratio of the scatterer size to the effective ``diffusion'' length.

To derive general scattering bounds, consider a 2D scatterer embedded in a possibly heterogeneous background. Passivity, which implies the absence of gain and that polarization currents do no work~\cite{Welters2014}, requires that the powers absorbed ($\Pabs$) and scattered ($\Pscat$) by the target body are non-negative~\cite{Miller2016}. These almost tautological conditions in fact dictate bounds on the largest currents that can be excited at the surface of any 2D structure. The key is that their sum, extinction ($\Pext = \Pabs + \Pscat$), is given by the imaginary part of a forward-scattering amplitude, which is a well-known consequence of the optical theorem~\cite{Newton1976,Jackson1999,Lytle2005}. For an arbitrarily shaped 2D scatterer with area $A$ that supports electric surface currents $\Kv$ (a magnetic-current generalization is given in the \SM), the absorbed and extinguished powers are given by~\cite{Jackson1999,Reid2015}
\begin{subequations}
\begin{align}
    \Pabs &= \frac{1}{2} \Re \int_A \cc{\Ev} \cdot \Kv \,{\rm d}A \label{eq:Abs} \\
    \Pext &= \frac{1}{2} \Re \int_A \cc{\Ev}_{\rm inc} \cdot \Kv \,{\rm d}A, \label{eq:Ext}
\end{align}
\label{eq:AbsExt}%
\end{subequations}
where, in the latter expression, $\int_A \cc{\Ev}_{\rm inc} \cdot \Kv \,{\rm d}A$ is a forward-scattering amplitude. A key feature of the optical theorem is that the extinction is the real part of an \emph{amplitude}, which is linear in the induced currents. By contrast, absorption is a \emph{quadratic} function of the currents/fields. Yet extinction must be \emph{greater} than absorption (due to the $\Pscat \geq 0$ condition noted above), requiring the linear functional to be greater than the quadratic one, a condition that cannot be satisfied for large enough currents. The inequality $\Pabs \leq \Pext$ thereby provides a convex constraint for \emph{any} optical-response function. Any optical-response maximization can thus be formulated as an optimization problem subject to this convex passivity constraint~\cite{Miller2016}. For a generic figure of merit $f(\Ev)$ of the fields (or, equivalently, currents), the design problem can be written
\begin{equation}
    \begin{aligned}
        & \text{maximize}  & & f(\Ev) \\
        & \text{subject to}  & & \Pabs(\Ev) \leq \Pext(\Ev).
    \end{aligned}
    \label{eq:MaxProb}
\end{equation}
Thanks to the convex nature of the constraint $\Pabs \leq \Pext$ and the simple expressions for $\Pabs$ and $\Pext$, \Eqref{MaxProb} can often be solved analytically---unlike the highly nonconvex Maxwell equations---thereby providing general upper-bound expressions without approximation.

To find bounds that solve \eqref{MaxProb}, we must specify a relationship between the field $\Ev$ and the induced current $\Kv$. To maintain generality we assume only that they are related by a linear operator $\Lke$,
\begin{align}
    \Lke \Kv = \Ev,    
    \label{eq:LKE}
\end{align}
where in different size, material, and parameter regimes, $\Lke$ may represent anything from a density-functional-theory operator~\cite{Yamamoto2006} or a hydrodynamic model~\cite{Mortensen2014,Raza2015}, to a simple scalar conductivity. For a scalar conductivity $\sigma$, $\Lke = 1/\sigma$. Given this current--field relation, the quadratic dependence of absorption on induced current, per \eqref{Abs}, is made explicit: $P_{\rm abs} = (1/2) \Re \int_A \cc{\Kv} \Lke \Kv \,{\rm d}A$. If we choose the figure of merit to be the absorbed or scattered power, then straightforward variational calculus (see \SM) from \eqref{MaxProb} yields the bounds
\begin{align}
    P_{\alpha} \leq \frac{1}{2} \beta_{\alpha} \int_A \cc{\Ev}_{\rm inc} \cdot \left(\Re \Lke\right)^{-1} \Ev_{\rm inc} \,{\rm d}A,
    \label{eq:GenPBounds}
\end{align}
where $\alpha$ denotes absorption, scattering, or extinction. The variable $\beta$ takes the values 
\begin{align}
    \beta_{\alpha} = \begin{cases}
        1 , \qquad & \alpha = {\rm absorption~or~extinction} \\
        \frac{1}{4}, \qquad &\alpha = {\rm scattering},
    \end{cases}
    \label{eq:betaAlpha}
\end{align}
which represent a power-balance asymmetry: absorption and extinction are maximized when $\Pabs = \Pext$, whereas scattering is maximized when $\Pscat = \Pabs = \Pext / 2$, akin to conjugate-matching conditions in circuit theory~\cite{Stutzman2012}. \Eqref{GenPBounds} sets a general bound, at any frequency, given only the incident field and the (material-driven) field--current relationship, dictated by the operator $\Lke$. The bounds apply in the far field, where $\Ev_{\rm inc}$ might be a plane wave or Bessel beam, as well as the near field, where $\Ev_{\rm inc}$ might be the field emanating from dipolar sources. Further below, we show that $\left(\Re \Lke\right)^{-1}$ can be considerably simplified in the case when $\Lke$ is the differential operator arising in nonlocal hydrodynamic models. First, however, we simplify \eqref{GenPBounds} for the important case of a spatially local conductivity.

A local conductivity $\sigmat$, relating currents at any point on the surface to fields at the same point, by $\Kv = \sigmat \Ev$, is the primary response model employed in the study of optical and plasmonic phenomena, in two as well as three dimensions. In 2D materials, it is common to have off-diagonal contributions to the conductivity (e.g. through magnetic-field biasing), and thus we allow $\sigmat$ to be a general $2\times2$ matrix (implicitly restricting $\Ev$ to its two components locally tangential to the 2D surface). Given that $\Lke = \sigmat^{-1}$, the term involving $\Lke$ in the bound of \eqref{GenPBounds} can be written: $\left(\Re \Lke\right)^{-1} = \sigmat^\dagger \left(\Re \sigmat\right)^{-1} \sigmat$. In far-field scattering, the quantity of interest is typically not the total absorbed or scattered power, but rather the cross-section, defined as the ratio of the power to the average beam intensity. The scattering cross-section, for example, is given by $\sigma_{\rm scat} = P_{\rm scat} / I_{\rm inc}$, where $I_{\rm inc} = |\Ev_{\rm inc}|^2_{\rm avg} / 2Z_0$. Then, the bound of \eqref{GenPBounds} simplifies for the absorption, scattering, and extinction cross-sections to
\begin{align}
    \frac{\sigma_{\alpha}}{A} \leq \beta_{\alpha} Z_0 \left\|\sigmat^\dagger \left(\Re \sigmat\right)^{-1} \sigmat\right\|_2
    \label{eq:PBounds}
\end{align}
where $Z_0$ is the impedance of free space, $\beta_\alpha$ is defined above in \eqref{betaAlpha}, and $\|\cdot\|_2$ denotes the induced matrix 2-norm~\cite{Trefethen1997} (which is the largest singular value of the matrix). The power of \eqref{PBounds} is its simplicity---the scattering efficiency of any 2D scatterer, whether it is a periodic array of circles~\cite{Thongrattanasiri2012}, a spherical coating~\cite{Christensen2015}, an isolated strip~\cite{DeAbajo2014}, or in any other configuration, has an upper bound defined solely by its material conductivity. We show below that simple ellipses can approach within $\approx10\%$ of the bounds, and that structures with two additional degrees of freedom can approach within $<1\%$ of the bounds. 

A key feature of the approach outlined here is that the optical response of a 2D material of interest can be cleanly delineated (without approximation) from the response of any ``background'' structures. Our formulation relies on the passivity constraints $\Pscat,\Pabs > 0$, and yet the choice of ``incident'' and ``scattered'' fields is arbitrary, as long as they sum to the total fields. As an example, there is significant interest in integrating 2D materials with photonic crystals~\cite{Zhan2012,Piper2014}; we can define the incident field that controls the bounds in \eqreftwo{GenPBounds}{PBounds} as the field in the presence of only the photonic crystal, and the scattered field as arising only from the addition of the 2D layer above it. The limits of \eqreftwo{GenPBounds}{PBounds}, as well as the limits derived below, then capture the maximum achievable enhancement due to the 2D material itself, subject to its inhomogeneous environment. Throughout this Letter we focus on free-standing graphene to understand its unique optical response, noting that generalization involving substrates and more complex surrounding structures can follow precisely this prescription.

Near-field optical response, in the presence of nearby emitters, is at least as important as far-field response. Here we find bounds to two important near-field quantities: (i) the local density of states (LDOS), which is a measure of the spontaneous-emission rate of a single excited dipole near the scatterer, and (ii) near-field radiative heat transfer, which is a measure of the radiation exchange between two bodies at different temperatures. The (electric) LDOS at a point $\xv$ is proportional to the power radiated by an (orientation-averaged) electric dipole at that point, and is given by the expression $\rho = (1/\pi \omega) \Im \sum_j \vect{p}_j \cdot \Ev_j(\xv)$, where $\Ev_j$ is the electric field excited by the dipole with moment $\vect{p}_j$, and where the sum over $j=x,y,z$ accounts for orientation-averaging~\cite{Novotny2012}. The expression for $\rho$ shows that LDOS is dictated by a causal amplitude (not a squared amplitude), exhibiting similar mathematical structure to extinction. The source of the similarity is that both extinction and LDOS can be decomposed into radiative and nonradiative components, which for the LDOS we denote by $\rho_{\rm rad}$ and $\rho_{\rm nr}$, respectively. The nonradiative part of the LDOS is given by the absorption in the scattering body (which is often an antenna), and per \eqref{Abs} is quadratic in the induced currents. Unlike far-field scattering, in the near field, the incident field increases rapidly at smaller distances $d$ ($|\vect{E}| \sim 1/d^3$). Thus, the same convex-optimization problem laid out in \eqref{MaxProb} leads to distance-dependent LDOS bounds via the replacements $\Pext \rightarrow \rho$ and $\Pabs \rightarrow \rho_{\rm nr}$. For an arbitrarily shaped 2D surface separated from the emitter by some minimum distance $d$, the bounds are (\SM):
\begin{align}
    \frac{\rho_{\alpha}}{\rho_0} \leq \frac{3\beta_{\alpha}}{8\left(k_0 d\right)^4} Z_0 \left\|\sigmat^\dagger \left(\Re \sigmat\right)^{-1} \sigmat\right\|_2
    \label{eq:LDOSBounds}
\end{align}
where $\alpha$ in this context denotes the total, radiative, or nonradiative component of the LDOS, $k_0 = \omega / c$, and $\rho_0$ is the free-space electric-dipole LDOS, $\rho_0 = \omega^2 / 2\pi^2 c^3$. Again $\beta_\alpha$ represents a power-balance (conjugate-matching) condition, and takes the value $1$ for nonradiative or total LDOS and $1/4$ for the radiative LDOS. \Eqref{LDOSBounds} includes the highest-order ($\sim1/d^3$) term from the incident electric field; lower-order terms ($\sim 1/d^2,1/d$) are generally negligible in the high-enhancement regimes of interest, as discussed quantitatively in \citeasnoun{Miller2016}. The $3/8$ coefficient in \eqref{LDOSBounds} is for the common case in which the surface is separated from the emitter by a separating plane; if the scattering body surrounds the emitter across a solid angle $\Omega$, the bound in \eqref{LDOSBounds} is multiplied by $4\Omega$. \Eqref{LDOSBounds} provides a general answer to the question of how efficient and effective a 2D optical antenna can be.

Radiative heat transfer (RHT), in which a warm body transfers energy to a colder one via photon exchange, is also subject to optical-response bounds. It has long been known~\cite{Polder1971,Rytov1988,Pendry1999} that near-field RHT can surpass the blackbody limit, as evanescent tunneling can outpace radiative exchange. Yet general limits to the process in conventional 3D materials had been unknown until our recent work~\cite{Miller2015}. The total RHT rate, $H$, is given by the net flux from one body at temperature $T_1$ to another at temperature $T_2$, typically expressed as~(\citeasnoun{Joulain2005}) $H_{1 \rightarrow 2} = \int_0^{\infty} \Phi(\omega) \left[\Theta(\omega,T_1) - \Theta(\omega,T_2)\right] \mathrm{d}\omega$, where $\Phi(\omega)$ is a temperature-independent energy flux and $\Theta$ is the Planck spectrum. The flux $\Phi$ is the power absorbed by the second body, having been emitted from the first, such that it is similar to the scattering problem bounded by \eqref{PBounds}. A key distinction is that the (incoherent) sources are in the \emph{interior} of one of the scattering bodies, invalidating the conventional optical theorem. This difficulty can be circumvented by breaking the flux transfer into two scattering problems, connected by a generalized~\cite{Kong1972} reciprocity relation (the material conductivity does \emph{not} need to be reciprocal), as outlined in \citeasnoun{Miller2015}. The key distinction in the case of 2D materials is the dimensionality of the domain over which the field intensities are evaluated, which for bodies with identical conductivities $\sigmat$ leads to the bound
\begin{align}
    \frac{\Phi}{\Phi_{\rm BB}} \leq \frac{3}{2\left(k_0 d\right)^4} Z_0^2 \left\|\sigmat^\dagger \left(\Re \sigmat\right)^{-1} \sigmat\right\|_2^2,
    \label{eq:RHTBounds}
\end{align}
where $d$ is the minimum separation distance between the arbitrarily shaped bodies, $\Phi_{\rm BB} = k_0^2 A / 4\pi^2$ is the blackbody limit (for infinite area $A$)~\cite{Joulain2005}, and the conductivity term is squared due to potential contributions from each body (see \SM). As for the LDOS bounds, \eqref{RHTBounds} assumes a separating plane between the bodies; corrugated surfaces that are interlaced (but non-touching) have bounds of the same functional form but with different numerical prefactors. An interesting 2D-specific aspect of \eqreftwo{LDOSBounds}{RHTBounds} is that they exhibit identical $1/d^4$ distance dependencies, whereas for 3D bodies, RHT increases more slowly for smaller separations ($\sim 1/d^2$) than does the LDOS ($\sim1/d^3$). 

\begin{figure}
    \centering\includegraphics[width=0.48\textwidth]{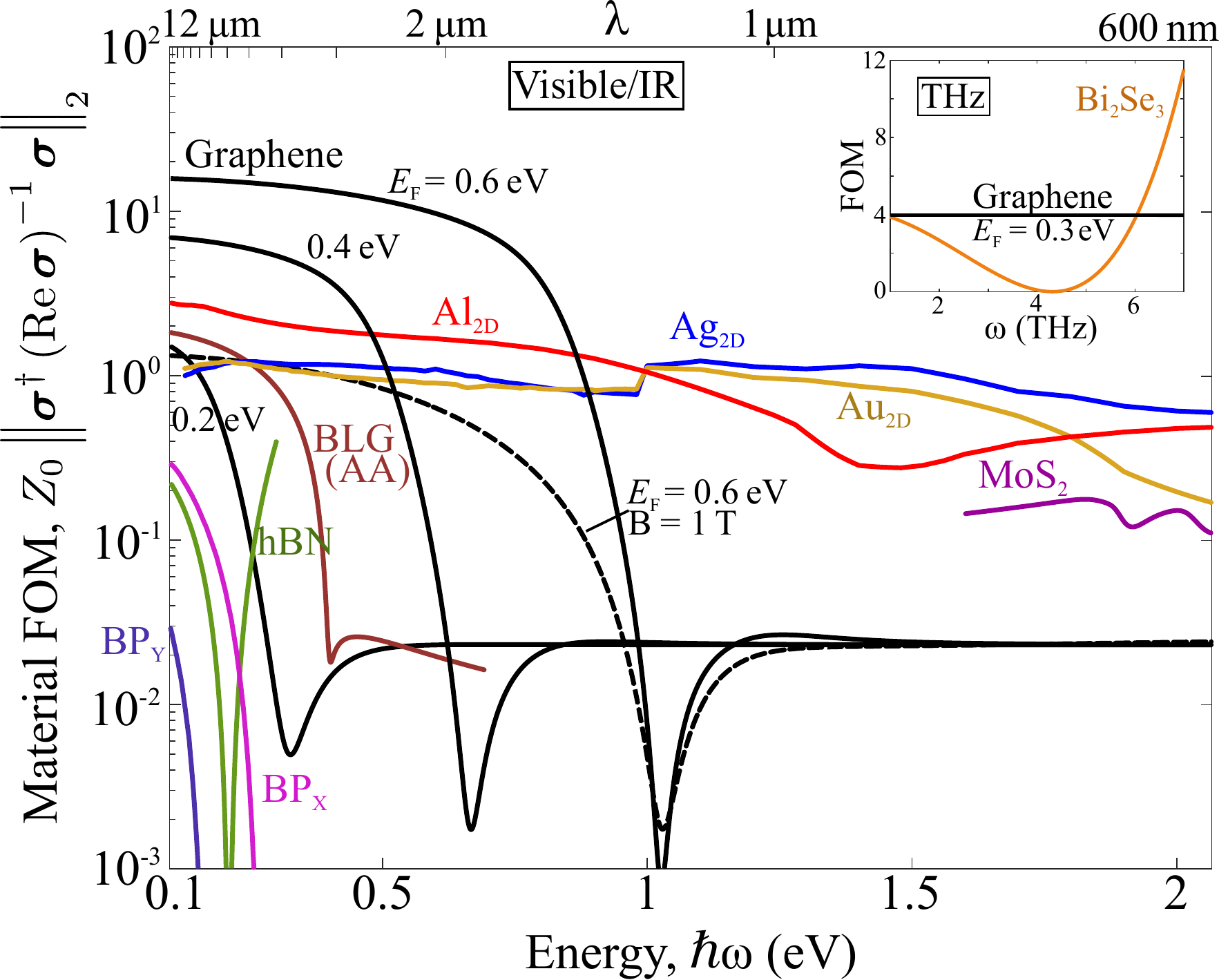}
    \caption{A simple material figure of merit (FOM), $Z_0 \left\|\sigmat^\dagger \left(\Re \sigmat\right)^{-1} \sigmat\right\|_2$ for conductivity $\sigmat$, dictates the maximum optical response that can be generated in 2D materials. Experimentally tabulated or analytically modeled optical data can be compared to assess optimal materials as they emerge. Here, we compare: graphene at different Fermi levels (solid black lines) and magnetic-biasing (dashed black line), AA-stacked bilayer graphene (dark red), hBN (green), \ce{MoS_2} (purple), the anisotropic conductivity components of black phosphorous (BP, pink and dark purple), and three 2D metals, Al (red), Ag (blue), and Au (gold). High-Fermi-level graphene and 2D silver  offer the largest possible responses at infrared and visible wavelengths, respectively. The inset compares graphene at THz frequencies to the topological insulator \ce{Bi2Se3}, which can have a surprisingly large FOM.}
    \label{fig:fig1}
\end{figure}
The fundamental limits of \eqrefrange{PBounds}{RHTBounds} share a common dimensionless material ``figure of merit'' (FOM), $Z_0 \big\|\sigmat^\dagger \left(\Re \sigmat\right)^{-1} \sigmat\big\|_2$. The FOM, which simplifies to $Z_0 |\sigma|^2 / \Re \sigma$ for a scalar conductivity, captures the intrinsic tradeoffs between high conductivity for large response and high losses that dissipate enhancement, and can be used to identify optimal materials. In \figref{fig1} we plot the FOM across a range of frequencies, using experimentally measured or analytically modeled material data for common 2D materials of interest: graphene, for various Fermi levels~\cite{Jablan2009}, magnetic biasing~\cite{Hanson2008}, and AA-type bilayer stacking~\cite{Wang2016} (at \SI{300}{\K}), hBN~\cite{Brar2014}, \ce{MoS_2}~\cite{Liu2014a}, black phosphourous (BP)~\cite{Low2016}, \ce{Bi2Se3} (at THz frequencies~\cite{DiPietro2013}), and metals Ag, Al, and Au, all taken to have 2D conductivities dictated by a combination~\cite{DeAbajo2015} of their bulk properties and their interlayer atomic spacing. Strongly doped graphene ($\EF = \SI{0.6}{\eV}$) offers the largest possible response across the infrared, whereas 2D Ag tends to be better in the visible. At THz frequencies, where graphene's potential is well-understood~\cite{Rana2008,Ju2011,Low2014}, the topological insulator \ce{Bi2Se3} shows promise for even larger response. More broadly, the simple material FOM, $|\sigma|^2 / \Re \sigma$ or its anistropic generalization $\left\|\sigmat^\dagger \left(\Re \sigmat\right)^{-1} \sigmat\right\|$, offers a metric for evaluating emerging (e.g. silicene~\cite{Vogt2012}, phosphorene~\cite{Xia2014,Liu2014}) and yet-to-be-discovered 2D materials. 

\begin{figure}
    \centering\includegraphics[width=0.48\textwidth]{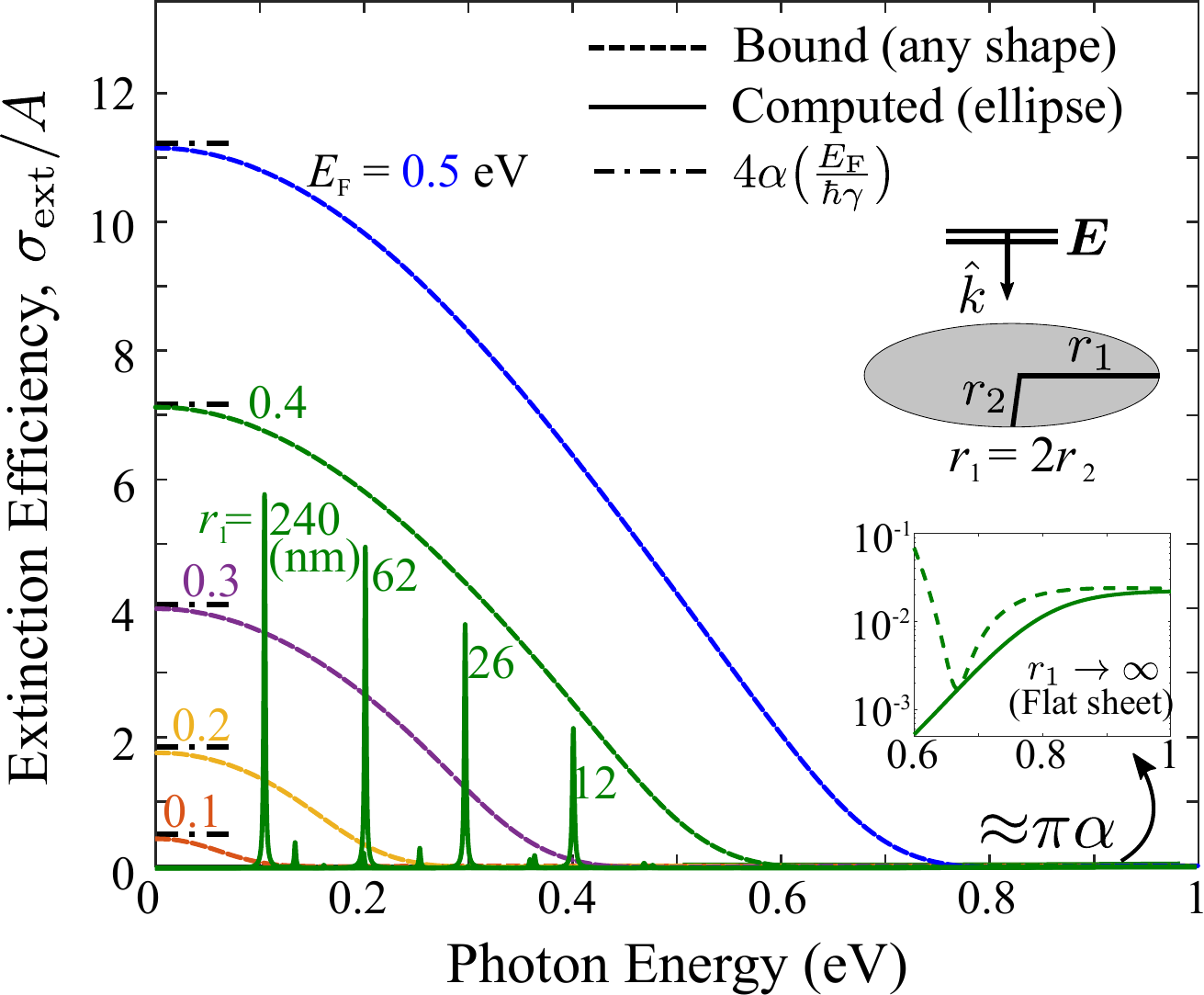}
    \caption{Upper limits (dashed lines) to the extinction cross-section of graphene scatterers of varying Fermi level, patterned into any shape, alongside the computed response of elliptical graphene disks of varying sizes for $\EF = \SI{0.4}{eV}$ (green, solid). The bounds, per \eqref{PBounds}, depend on graphene's 2D conductivity and incorporate the extent to which losses can be overcome. The disks reach within $\approx 85\%$ of the bounds, and in the {\SM} we show that slightly more exotic shapes can reach $>99\%$ of the bounds. Simple asymptotic expressions for the bounds emerge at low (dash--dot lines) and high frequencies. In the high-frequency limit, the limits converge to $\pi\alpha$, and are thereby reached with a simple flat sheet (inset).}
    \label{fig:fig2}
\end{figure}

In the following we specialize our considerations to graphene, the standard-bearer for 2D materials, to examine the degree to which the bounds of \eqrefrange{PBounds}{RHTBounds} can be attained in specific structures. We adopt the conventional local description, including intra- and interband dispersion. Appropriate modifications~\cite{DeAbajo2014,Jablan2009} are included to account for a finite intrinsic damping rate, $\gamma = 1/\tau = \left(\SI{e12}{\eV/\second}\right)/\EF$, which is taken as Fermi-level-dependent (corresponding to a Fermi-level-independent mobility), with a magnitude mirroring that adopted in \citeasnoun{DeAbajo2014}. \Figref{fig2} shows the cross-section bounds (dashed lines), per \eqref{PBounds}, alongside graphene disks (with $\EF = \SI{0.4}{eV}$) that approach the bounds at frequencies across the infrared. For simplicity, we fix the aspect ratio of the disks at 2:1 and simply reduce their size to increase their resonant frequency; each disk achieves $\approx 85\%$ of its extinction cross-section bound. The disk diameters are kept greater than $\SI{10}{nm}$ to ensure the validity of our local description. We employ a fast quasistatic solver~\cite{Christensen2017} to compute the response of the ellipses, which is verified with a free-software implementation~\cite{ReidScuffEM} of the boundary element method (BEM)~\cite{Harrington1993} for the full electrodynamic problem with the surface conductivity incorporated as a modified boundary condition~\cite{Reid2015}. If edge scattering, or any other defect, were to increase the damping rate, such an increase could be seamlessly incorporated in the bounds of \eqrefrange{PBounds}{RHTBounds} through direct modification of the conductivity. In the \SM, we show that with two extra geometrical degrees of freedom (e.g., a ``pinched ellipse''), one can reach $>99.6\%$ of the bound. The cross-section bounds can also be used as bounds on the fill fraction of graphene required for perfect absorption in a planar arrangement, and they suggest the potential for an order-of-magnitude reduction relative to the best known results~\cite{Thongrattanasiri2012}. Conversely, such room for improvement could be used to significantly increase the perfect-absorption bandwidth beyond the modern state-of-the-art.

The bounds simplify analytically at the low- and high-frequency extremes. In these regimes, graphene's isotropic conductivity is real-valued and comprises simple material and fundamental constants, such that the material FOM is approximately 
\begin{align}
    Z_0 \left\| \sigmat^\dagger \left(\Re \sigmat\right)^{-1} \sigmat \right\|_2
    \approx Z_0 \sigma \approx
    \begin{cases}
        4\alpha \big(\frac{\EF}{\hbar \gamma}\big) & \omega \ll \gamma \\ %
        \pi\alpha & \omega \gg 2\EF / \hbar.
    \end{cases}
    \label{eq:GrapheneApprox}
\end{align}
The low-frequency proportionality to $\EF/\hbar\gamma$ arises as a consequence of the \emph{intra}band contributions to the conductivity, in contrast to the \emph{inter}band dominance at high frequencies. Interband contributions to the conductivity are often ignored at energies below the Fermi level, but even at those energies they are responsible for a sizable fraction of the loss rate, thus causing the quadratic roll-off (derived in \SM) of the maximum efficiency seen on the left-hand side of \figref{fig2}.

Famously, at high frequencies a uniform sheet of graphene has a scattering efficiency $\sigma/A \approx \pi\alpha$ (Refs.~\cite{Nair2008,Kuzmenko2008,Mak2008}). Interestingly, \figref{fig2} and \eqref{GrapheneApprox} reveal that $\pi\alpha$ is the \emph{largest possible} scattering efficiency, for \emph{any} shape or configuration of graphene, at those frequencies. Per the incident-field discussion above, it is possible to increase the absolute absorption of a plane wave at those frequencies by structuring the background (e.g. with a photonic-crystal slab supporting the graphene), but the percentage of the background field intensity that can be absorbed by the graphene is necessarily $\leq \pi\alpha$, no matter how the graphene is structured. The right-hand side of \figref{fig2} shows the bounds for each Fermi level converging to $\pi\alpha$, with the inset magnifying the high-energy region and showing that the response of a flat sheet indeed reaches the bound.

\begin{figure}
    \centering\includegraphics[width=0.48\textwidth]{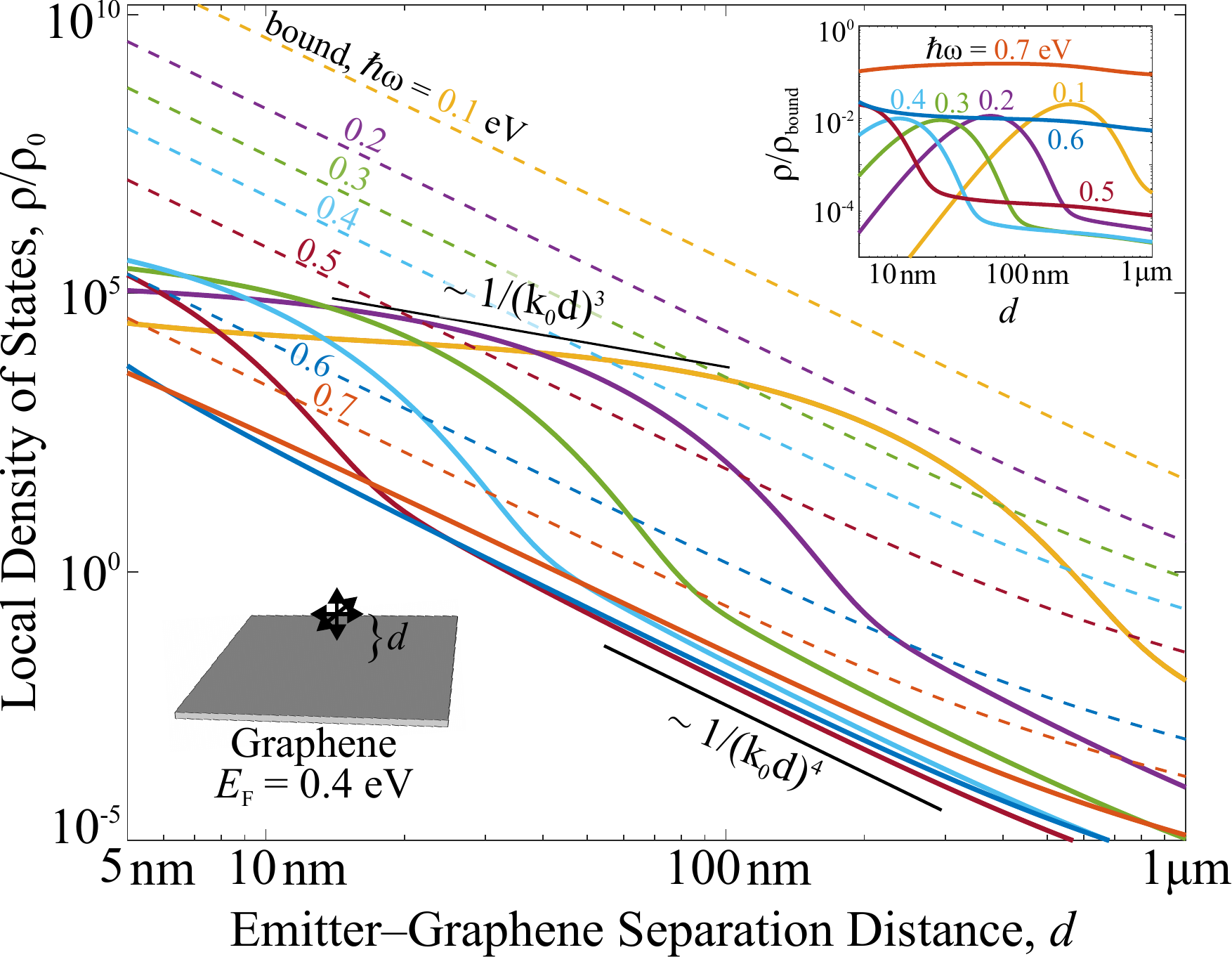}
    \caption{Comparison of the LDOS above a flat graphene sheet (dashed lines) to the LDOS bounds for any structure (solid lines), for multiple frequencies (colored lines) and as a function of the emitter--graphene separation distance $d$ (with $\EF = \SI{0.4}{\eV}$). For larger separations and higher frequencies, the LDOS above a flat sheet follows the ideal $\sim1/d^4$ scaling, but at shorter separations and lower frequencies (where the response is potentially largest), the optimal-frequency response follows a $\sim1/d^3$ envelope. The inset shows of ratio of the flat-sheet LDOS to the upper bound, showing that there is the potential for 1--2 orders of magnitude improvement.}
    \label{fig:fig3}
\end{figure}
The near-field LDOS and RHT limits are more challenging to attain. We study the LDOS near a flat sheet of graphene, the most common 2D platform for spontaneous-emission enhancements to date~\cite{Koppens2011,Gaudreau2013,Tielrooij2015}, and show that there is a large performance gap between the flat-sheet response and the fundamental limits of \eqref{LDOSBounds}. There are two key factors that control the near-field bounds (for both LDOS and RHT): the material FOM $|\sigma|^2 / \Re \sigma$, and a ``near-field enhancement factor'' $1/d^4$, for emitter--sheet distance $d$. The $1/d^4$ near-field enhancement factor is particularly interesting, because it increases more rapidly than in 3D materials (for which the LDOS~\cite{Miller2016} and RHT~\cite{Miller2015} bounds scale as $1/d^3$ and $1/d^2$, resp.). In \figref{fig3}, we show the LDOS as a function of the emitter--graphene separation, for a fixed Fermi level $\EF = \SI{0.4}{\eV}$ and a range of frequencies (colored solid lines). The bounds for each frequency are shown in the colored dashed lines, and the ratio of the LDOS $\rho$ to the LDOS bound $\rho_{\rm bound}$ is shown in the inset. For low and moderate frequencies, there is an ideal distance at which the LDOS most closely approaches its frequency-dependent bound, whereas the high-frequency regime (e.g. $\hbar\omega = \SI{0.7}{eV}$) is almost distance-insensitive due to high losses. 

\Figref{fig3} shows two asymptotic distance-scaling trends. First, at high frequencies and/or large separations (\SI{50}{\nm} to \SI{1}{\um}), the LDOS enhancement scales as $1/(k_0d)^4$. We show in the {\SM} that in this regime the LDOS further exhibits the material-enhancement factor $|\sigma|^2/\Re \sigma$, falling short of the bound only by a factor of 2. In this regime, the LDOS is dominated by a ``lossy-background'' contribution~\cite{Gaudreau2013}, which is insensitive to details of the plasmonic mode, and due instead predominantly to interband absorption in graphene (permitted even below $2\EF$ for nonzero temperatures). Of more interest may be the opposite regime---higher frequencies at smaller separations---which are known~\cite{Rodriguez-Lopez2015} to have reduced distance dependencies. It is crucial to note that the bounds presented in this Letter are \emph{not} scaling laws; instead, at each frequency and distance they represent independent response limits. We see in \figref{fig3} that for each individual frequency, $\rho/\rho_0$ flattens towards a constant value at very small distances, because the corresponding plasmon surface-parallel wavenumber is smaller than $1/d$ and does not change; however, the \emph{envelope} formed over many frequencies (for a given separation $d$) shows a $1/(k_0 d)^3$ as higher-wavenumber plasmons are accessed at smaller distances. This suggests a simple potential approach to reach the bound: instead of finding a geometrical configuration that approaches the bound at all frequencies and separations, concentrate on finding a structure that reaches the bound at a single frequency and separation of interest. A ``family'' of structures that combine to approach the bounds over a large parameter regime may then naturally emerge.

\begin{figure}
    \centering\includegraphics[width=0.48\textwidth]{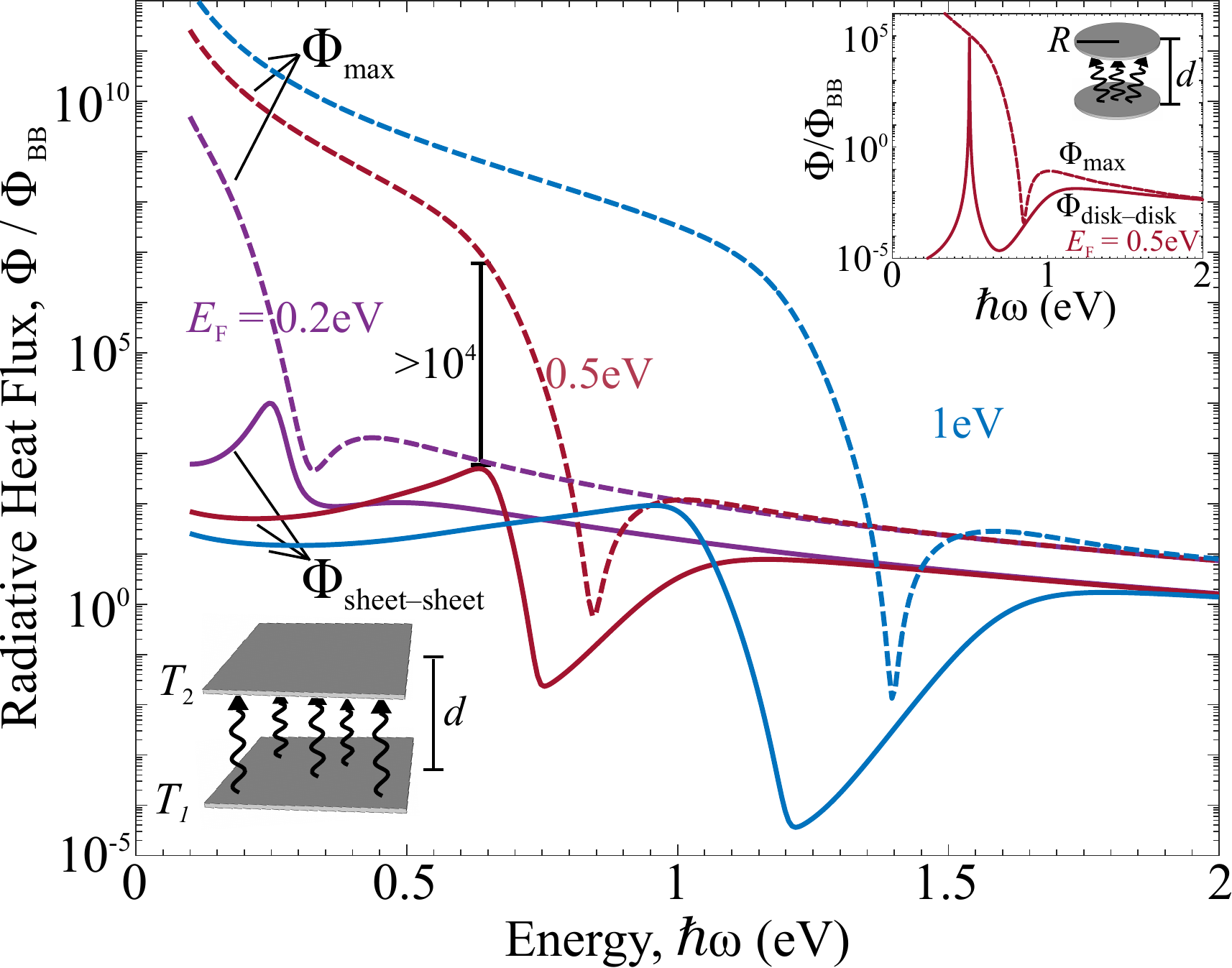}
    \caption{Radiative heat flux between two graphene structures (at $T=\SI{300}{\K}$ and $d=\SI{10}{\nm}$), for flat sheets (solid lines) and for the arbitrary-shape analytical bounds (dotted lines). For a Fermi level $\SI{0.5}{\eV}$, the flat sheets fall short of the bounds by $10^4$ at their peak, due to near-field interference effects between the sheets. The interference effects do not arise between dipolar circles (inset), whereby the bound is nearly achieved (for $R=\SI{5}{\nm}$ and $d=\SI{30}{\nm}$). The discrepancy between the disk and flat-sheet RHT rates suggests the possibility of significant improvement via patterning.}
    \label{fig:fig4}
\end{figure}
Near-field RHT shows similar characteristics, in which the bounds may be approached with flat graphene sheets at specific energy, Fermi-level, and separation-distance parameter combinations. As a counterpart to the LDOS representation of \figref{fig3}, in \figref{fig4} we fix the separation distance at $\SI{10}{\nm}$ and plot the frequency-dependent RHT~\cite{Ilic2012} for three Fermi levels. The respective bounds, from \eqref{RHTBounds}, show the same ``dip'' as seen in the inset of \figref{fig2}(b), which occurs at the frequency where the imaginary part of the conductivity crosses zero. At these frequencies, the RHT between flat sheets can approach the bounds. However, at other frequencies, where the potential RHT is significantly larger, the flat sheets fall short by orders of magnitude, as depicted in \figref{fig4} at $\EF = \SI{0.5}{\eV}$. The flat-sheet case falls short due to near-field interference effects: as the sheets approach each other, the plasmonic modes at each interface interact with each other, creating a level-splitting effect that reduces their maximum transmission to only a narrow band of wavevectors~\cite{Miller2016}. By contrast, for two dipolar circles in a quasistatic approximation (\figref{fig4} inset), the RHT between the two bodies can approach its respective bound. These examples suggest that patterned graphene sheets, designed to control and optimize their two-body interference patterns, represent a promising approach towards reaching the bounds and thereby unprecedented levels of radiative heat transfer. In the \SM, we show that achieving RHT at the level of the bound, even over the narrow bandwidths associated with plasmonic resonances, would enable radiative transfer to be greater than conductive transfer through air at separations of almost $\SI{1}{\mu m}$, significantly larger than is currently possible~\cite{Miller2016}.

Having examined the response of graphene structures in the local-conductivity approximation, we now reconsider \emph{nonlocal} conductivity models. For structures in the \SIrange{2}{10}{\nm} size range, below the local-conductivity regime but large enough to not necessitate fully quantum-mechanical models, hydrodynamic conductivity equations~\cite{Ciraci2012,Mortensen2014,Raza2015}, or similar gradient-based models of nonlocality~\cite{Hanson2008,Fallahi2015}, can provide an improved account of the optical response. In a hydrodynamic model, the currents behave akin to fluids with a diffusion constant $D$ and convection constant $\beta$ (both real-valued), with a current--field relation given by~\cite{Mortensen2014}
\begin{align}
    \underbrace{\left[\frac{-i}{\epsilon_0 \omega \omegap^2} \left(\beta^2 + D\left(\gamma - i\omega \right)\right) \nabla\nabla \cdot + \sigma_{\rm loc}^{-1}\right]}_{\Lke} \Kv = \Ev,
    \label{eq:HydroEqn}
\end{align}
where $\sigma_{\rm loc}$, $\omegap$, and $\gamma$ are the local conductivity, plasma frequency, and damping rate of the 2D material, respectively. Per \eqref{GenPBounds}, the 2D-material response bounds depend only on the Hermitian part of the $\Lke$ operator, denoted by an underbrace in \eqref{HydroEqn}. Before deriving bounds dependent on the hydrodynamic parameters, we note that the grad--div  hydrodynamic term in \eqref{HydroEqn} cannot increase the maximum optical response. The operator $-\nabla\nabla \cdot$ is a positive semidefinite Hermitian operator (for the usual $L^2$-space overlap-integral inner product), shown by integration by parts in conjunction with the no-spillout boundary condition. The Hermiticity of the grad--div operator means that the Hermitian part of $\Lke$ is given by $\Re \Lke = (\Lke + \Lke^{\dagger})/2 = -\frac{D}{\omegap^2} \nabla \nabla \cdot + \Re \sigma_{\rm loc}^{-1}$. Because $-\nabla \nabla \cdot$ is a positive-semidefinite addition to the positive-semidefinite term $\Re \sigma_{\rm loc}^{-1}$, $\left\|\left(\Re \Lke\right)^{-1}\right\| \leq \left(\Re \sigma_{\rm loc}^{-1}\right)^{-1} = |\sigma_{\rm loc}|^2 / \Re \sigma_{\rm loc}$. Thus the nonlocal response is subject to the bound imposed by the underlying local conductivity, demonstrating that nonlocal effects of this type \emph{cannot} surpass the local-conductivity response explored in depth above. 

We can further show that hydrodynamic nonlocality necessarily \emph{reduces} the maximum achievable optical response in a given 2D material, by exploiting the quasistatic nature of electromagnetic interactions at the length scales for which nonlocal effects manifest. The key insight required to derive bounds subject to the nonlocal current--field relation, \eqref{HydroEqn}, is that the absorbed power can be written as a quadratic form of both the currents $\Kv$ as well as $\nabla \cdot \Kv$ (proportional to the induced charge): $\Pabs = (1/2) \Re \int_A \cc{\Kv} \cdot \Ev = 1/2 \int_A \left[a (\nabla \cdot \cc{\Kv})(\nabla \cdot \Kv) + b\cc{\Kv}\cdot \Kv\right]$, where $a=D/\omegap^2$ and $b=\Re(\sigma_{\rm loc}^{-1})$. Similarly, the extinction can be written as a \emph{linear} function of either $\Kv$ or $\nabla \cdot \Kv$ (exploiting the quasistatic nature of the fields), such that $\Pabs \leq \Pext$ offers two convex constraints for the generalized nonlocal-conductivity problem. We defer to the {\SM} for a detailed derivation of general figures of merit under this constraint, and state a simplified version of the result for the extinction cross-section. The additional $\nabla \cdot \Kv$ constraint introduces a size dependence in the bound, in the form of a ``radius'' $r$ that is the smallest bounding sphere of the scatterer along the direction of the incident-field polarization. Defining a plasmonic ``diffusion'' length $\ell_D = \sqrt{cD/\omegap^2}$ (for speed of light $c$), the variational-calculus approach outlined above yields an analogue of \eqref{PBounds} in the presence of a hydrodynamic nonlocality:
\begin{align}
    \frac{\sigma_{\rm ext}}{A} \leq \left[ \left(Z_0 \frac{\left| \sigma_{\rm loc} \right|^2}{\Re \sigma_{\rm loc}}\right)^{-1} +  \left(\frac{r^2}{\ell_D^2}\right)^{-1} \right]^{-1}.
    \label{eq:nlBnd}
\end{align}
\Eqref{nlBnd} has an appealing, intuitive interpretation: the cross-section of a scatterer is bounded above by a combination of the local-conductivity bound and a nonlocal contribution proportional to the square of the ratio of the size of the scatterer to the ``diffusion'' length. Thus as the size of the particle approaches $\ell_D$, and goes below it, there must be a significant reduction in the maximal attainable optical response. There is ambiguity as to what the exact value of $D$, or equivalently $\ell_D$, should be in 2D materials such as graphene; the bounds developed serve as an impetus for future measurement or simulation, to delineate the sizes at which the local/non-local transition occurs. Conversely, since the bound shows a dramatic reduction at sizes below $\ell_D$, \eqref{nlBnd} can serve as a means to extract this nonlocal property of any 2D material from experimental measurements.

General limits serve to contextualize a large design space, pointing towards phenomena and performance levels that may be possible, and clarifying basic limiting factors. Here we have presented a set of optical-response bounds for 2D materials, generalizing recent 3D-material bounds~\cite{Miller2015,Miller2016} to incorporate both local and nonlocal models of 2D conductivities. We further studied the response of standard graphene structures---ellipses and sheets---relative to their respective bounds, showing that the far-field absorption efficiency bounds can be reliably approached within $10\%$, but that the near-field bounds are approached only in specific parameter regimes, suggesting the possibility for design to enable new levels of response. The figure of merit $\left\|\sigmat^\dagger \left(\Re \sigmat\right)^{-1} \sigmat\right\|$ can serve to evaluate new 2D materials as they are discovered, and their optical properties are measured. Our results point to a few directions where future work may further clarify the landscape for 2D-material optics. One topic of current interest is in patterned gain and loss~\cite{Pick2017,Cerjan2016a} (esp. $\mathcal{PT}$-symmetry~\cite{Guo2009,Regensburger2012,Cerjan2016}), which exhibit a variety of novel behaviors, from exceptional points to loss-induced transparency. Our bounds depend on passivity, which excludes gain materials, but in fact the bounds only require passivity \emph{on average}, i.e., averaged over the structure. Thus \eqrefrange{GenPBounds}{RHTBounds} should be extensible to patterned gain--loss structures. A second area for future work is in exploration of quantum models of the $\Lke$ operator. We have shown here explicit bounds for the cases of local and hydrodynamic conductivities, but there is also significant interest in quantum descriptions of the response. Through, for example, density-functional theory~\cite{Burke2005}, analytical bounds in such cases may lead to a continuum of optical-response limits across classical, semi-classical, and quantum regimes.


\begin{acknowledgements} 
    O.D.M. was supported by the Air Force Office of Scientific Research under award number FA9550-17-1-0093. O.I. and H.A.A. were supported as part of the DOE ``Light-Material Interactions in Energy Conversion’’ Energy Frontier Research Center under grant DE-SC0001293, and acknowledge support from the Northrop Grumman Corporation through NG Next. T.C. was supported by the Danish Council for Independent Research (grant no. DFFC6108-00667). M.S. was partly supported (reading and analysis of the manuscript) by S3TEC, an Energy Frontier Research Center funded by the U.S. Department of Energy under grant no. DE-SC0001299. J.D.J., M.S., and S.G.J. were partly supported by the Army Research Office through the Institute for Soldier Nanotechnologies under contract no. W911NF-13-D-0001.
\end{acknowledgements}

\bibliography{/Users/odm5/Library/texmf/bibtex/bib/library,/Users/odm5/Library/texmf/bibtex/bib/journalshort,/Users/odm5/Library/texmf/bibtex/bib/nonotes,/Users/odm5/Library/texmf/bibtex/bib/My_Pubs_abbr}

\end{document}



\title{Supporting Information: Limits to the Optical Response of Graphene and 2D Materials}

\author{Owen~D.~Miller}
\email[Corresponding author: ]{owen.miller@yale.edu} 
\affiliation{Department of Applied Physics, Yale University, New Haven, CT 06511}
\author{Ognjen~Ilic}
\affiliation{Department of Applied Physics and Material Science, California Institute of Technology, Pasadena, CA 91125}
\author{Thomas~Christensen}
\affiliation{Department of Physics, Massachusetts Institute of Technology, Cambridge, MA 02139}
\author{M.~T.~Homer Reid}
\affiliation{Department of Mathematics, Massachusetts Institute of Technology, Cambridge, MA 02139}
\author{Harry~A.~Atwater}
\affiliation{Department of Applied Physics and Material Science, California Institute of Technology, Pasadena, CA 91125}
\author{John~D.~Joannopoulos}
\affiliation{Department of Physics, Massachusetts Institute of Technology, Cambridge, MA 02139}
\author{Marin~Solja\v{c}i\'{c}}
\affiliation{Department of Physics, Massachusetts Institute of Technology, Cambridge, MA 02139}
\author{Steven~G.~Johnson}
\affiliation{Department of Physics, Massachusetts Institute of Technology, Cambridge, MA 02139}
\affiliation{Department of Mathematics, Massachusetts Institute of Technology, Cambridge, MA 02139}



\maketitle

\tableofcontents

\sisetup{range-phrase=--}
\sisetup{range-units=single}

\section{Optimized structure to reach within $1\%$ of extinction bound}
\begin{figure}
    \centering\includegraphics[width=0.48\textwidth]{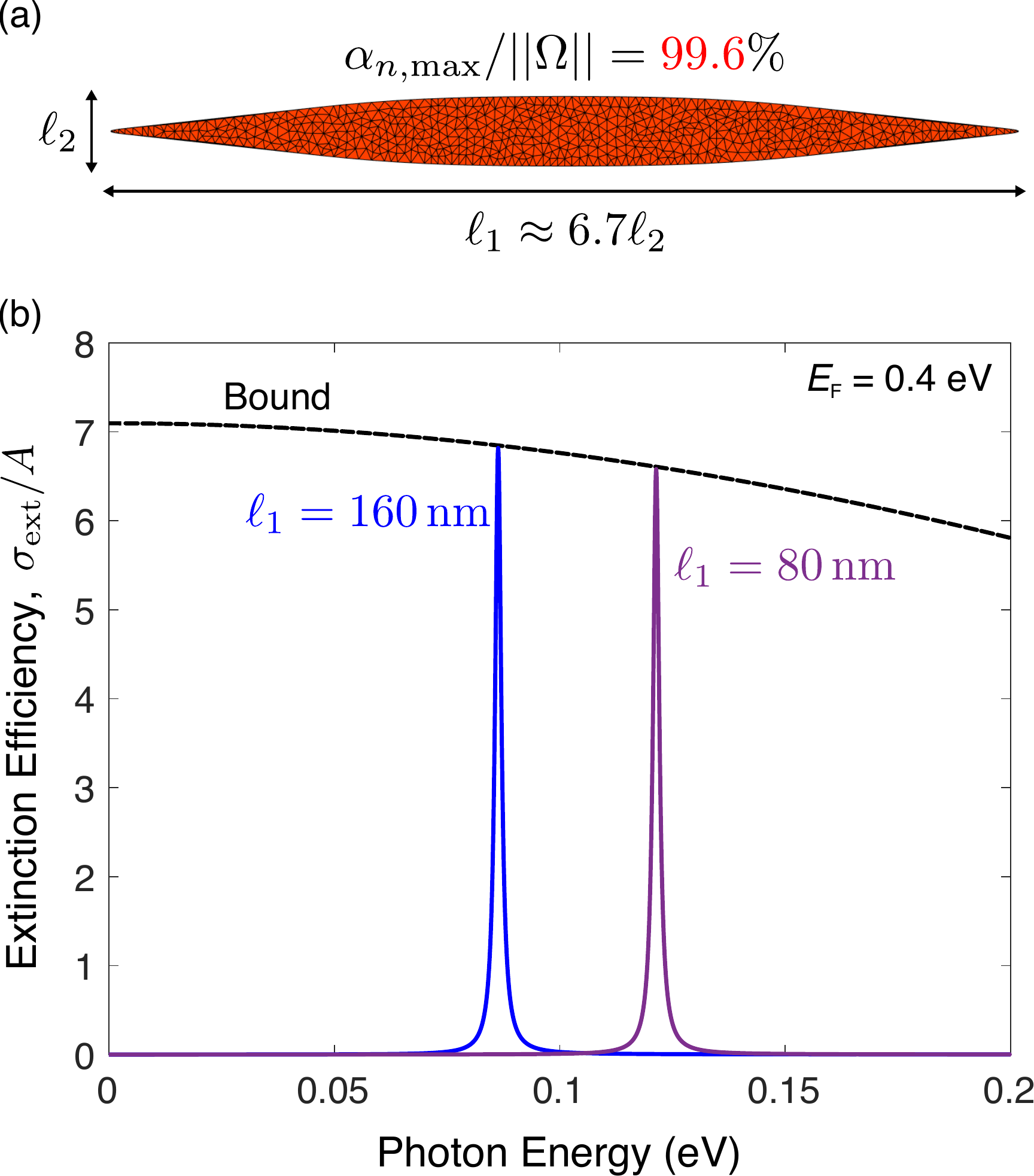}
    \caption{(a) ``Pinched ellipse'' geometry, described by \eqref{parameterization}, with the parameters in \eqref{paramvals}. The pinched ellipse geometry has a mode with $99.6\%$ of the maximum polarizability possible, such that the response is almost perfectly concentrated at a single resonant frequency. (b) Spectral response of the pinched-ellipse geometry, for two different scaling factors (given by the widths of the structures). The response achieves $99.6\%$ of the general bound.}
    \label{fig:figS1}%
\end{figure}
In this section we show that the bounds can be reached to within $1\%$ through simple optimization of the scattering structure. The elliptical disks considered in the main text only have two degrees of freedom, one of which is a scaling parameter that solely shifts the frequency. Thus, we consider the ``pinched ellipse'' structure depicted in \figref{figS1}. Utilizing the angle $\theta$ in the two-dimensional plane of the structure, the boundary of a simple ellipse can be parameterized as $x = a\cos\theta$, $y=\sin\theta$. We generate the pinched ellipse via the parameterization:
\begin{subequations}
\begin{align}
    x &= a \cos \theta \\
    y &= \sin \theta \left[ 1 + d e^{-|x(\theta)|^s/w} \right] 
\end{align}
    \label{eq:parameterization}%
\end{subequations}
where $a$, $d$, $s$, and $w$ are free parameters. Many different combinations can lead to good performance; from simple unconstrained optimizations, the values
\begin{subequations}
\begin{align}
    a &= 53.788 \\
    d &= 3.0917 \\
    s &= 3.6358 \\
    w &= 0.3964
\end{align} \label{eq:paramvals}%
\end{subequations}
reach near-ideal performance. The performance of such a structure is exhibited not only in the peak of the spectral response but also in the quasistatic polarizability. The quasistatic polarizability of a 2D scatterer, $\alpha(\omega)$, can be decomposed into a complete set of modes that are orthonormal under a properly chosen inner product. The polarizabilities of the modes, $\alpha_n$ for mode $n$, must satisfy the sum rule~\cite{DeAbajo2015}
\begin{align}
    \sum_n \alpha_n \leq ||\Omega||
\end{align}
where $||\Omega||$ is the total surface area of the scatterer. 

The capability of a structure to reach the bounds developed in the main text is directly related to whether its response is concentrated into a single mode at the frequency of interest. The elliptical disks of the main text have oscillator strengths, i.e., mode polarizabilities, of approximately $90\%$, explaining their large extinction cross-sections that reach within 10\% of the bounds. For the pinched ellipse of \figref{figS1}, the parameter values in \eqref{paramvals} yield a normalized oscillator strength of $99.6\%$, as computed by a quasistatic integral-equation solver~\cite{Christensen2017} and shown in \figref{figS1}(a). Such a large oscillation strength indicates that the scatterer should reach $99.6\%$ of the extinction bound, which we verify numerically. The nearly ideal spectral response is shown in \figref{figS1}(b), for two scaled versions of the ellipse shown in \figref{figS1}(a) with the parameters given in \eqref{paramvals}.

\section{Optimal conductive heat transfer through graphene}
\begin{figure}
    \centering\includegraphics[width=0.48\textwidth]{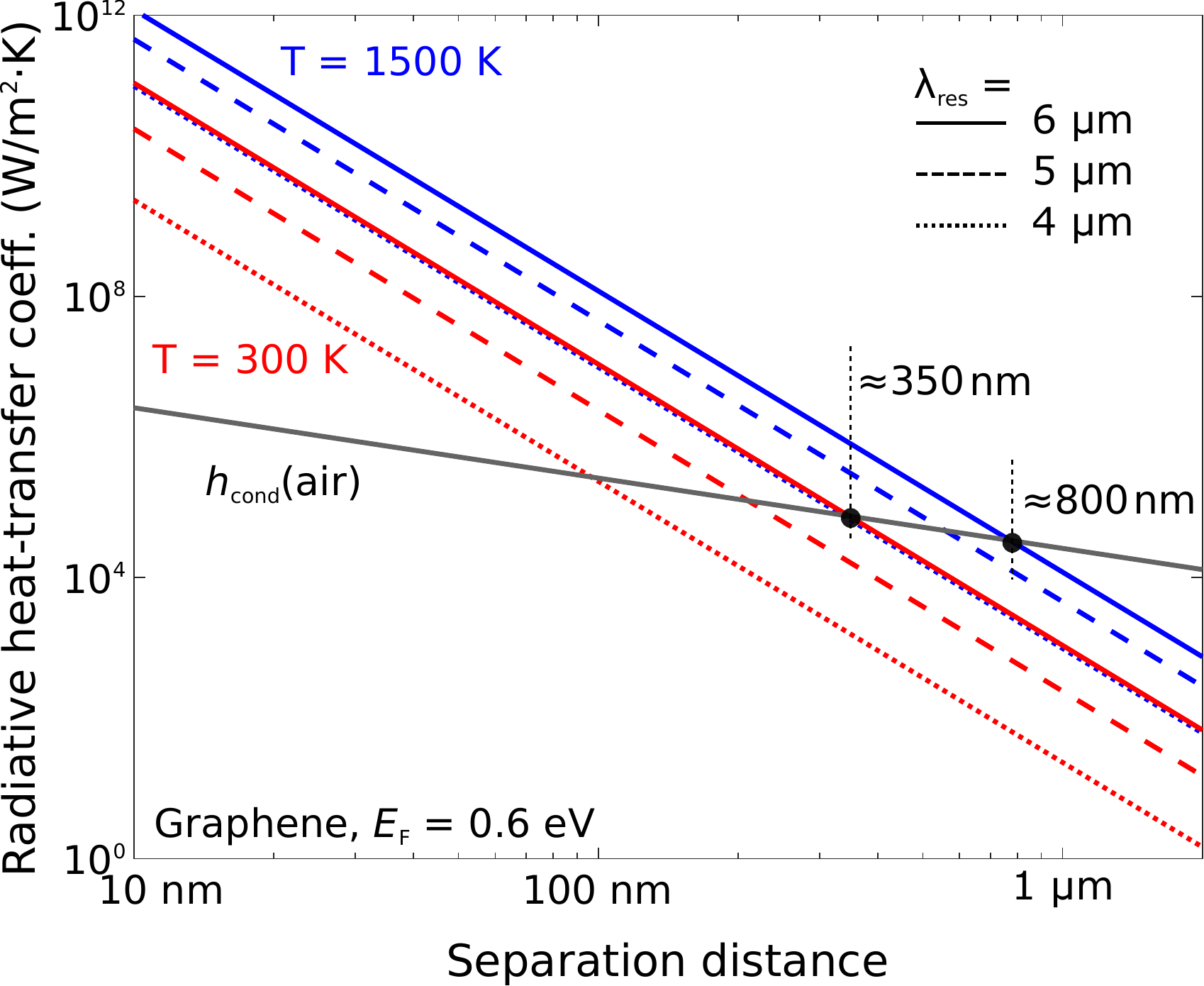}
    \caption{Optimal radiative heat-transfer coefficient for near-field energy exchange between graphene structures operating at the maximum theoretical flux rate, over a bandwidth dictated by the material loss rate. At $\SI{300}{K}$ it is possible for graphene RHT to surpass conductive transfer through air at $\approx \SI{350}{nm}$ separation distance; at $\SI{1500}{K}$, it is possible at almost $\SI{800}{nm}$ separations. The theoretical RHT coefficient increases with the resonant wavelength, $\lres$, due to the increasing material FOM $|\sigma|^2 / \Re \sigma$ of graphene with increasing wavelength.}
    \label{fig:figS2}
\end{figure}
We showed in Eq.~(8) of the main text that near-field radiative heat transfer (RHT) has a unique $1/d^4$ separation-distance dependence for 2D materials, increasing more rapidly than the $1/d^3$ dependence of 3D materials. Here we consider the potential for a 2D material such as graphene to exhibit large radiative heat transfer relative to the large \emph{conductive} heat transfer rate for two bodies separated by micron-scale air gaps. As discussed in the main text, the total radiative heat transfer between two bodies is given by $H = \int \Phi(\omega) \left[\Theta(\omega,T_1) - \Theta(\omega,T_2)\right]\,{\rm d}\omega$. For a small temperature differential between the bodies, the \emph{conductance} (heat transfer per unit temperature) per area $A$ is termed the \emph{radiative heat transfer coefficient} and is given by 
\begin{align} 
    h_\text{rad} &= \frac{1}{A} \int \Phi(\omega) \frac{\partial \Theta}{\partial T} \,{\rm d}\omega = \frac{1}{A} k_B \int \Phi(\omega) f(\omega) \,{\rm d}\omega, 
\end{align} 
where 
\begin{align} 
    f(\omega) = \left(\frac{\hbar \omega}{k_B T}\right)^2 \frac{e^{\hbar\omega/k_BT}}{\left(e^{\hbar\omega/k_BT} - 1\right)^2} 
\end{align}
For common 2D materials such as graphene, the material loss rates are sufficiently small that resonant response is sharply peaked, with a width determined by the material loss. For resonant response the distribution of $\Phi(\omega)$ will be much sharper than the Boltmann-like distribution $f(\omega)$ in the integrand. Thus we can approximate $h$ by 
\begin{align} 
    h_\text{rad} &\approx \frac{1}{A} k_B f(\wres) \int \Phi(\omega) \,{\rm d}\omega \\
                 &\approx \frac{1}{A} k_B f(\wres) \Phi(\wres) \frac{\pi \Delta\omega}{2}
                 \label{eq:hrad_eqn}
\end{align}
where $\wres$ is the peak resonant frequency, and the second approximation assumed a Lorentzian distribution for $\Phi$, with $\Delta \omega$ as the full-width at half-maximum of the distribution. For a plasmonic material such as graphene, we can model the bandwidth through the quality factor: $Q = \frac{\omega}{\Delta \omega} = \frac{|\Im \sigma|}{\Re \sigma}$, which is the 2D-material version of the well-known expression $Q = |\Re \chi| / \Im \chi$ (Refs.~\cite{Wang2006a,Raman2013}). For graphene and similar materials at optical frequencies, $\Im \sigma \approx |\sigma|$. Thus if we use the minimal material-dependent bandwidth $\Delta \omega \approx \wres \Re \sigma / |\sigma|$, and insert the bound for $\Phi/A$ from Eq.~(8) in the main text into \eqref{hrad_eqn}, we find a bound on the radiative heat-transfer coefficient:
\begin{align}
    h_{\rm rad} \leq \frac{3}{16\pi^2} \frac{k_B \wres}{d^2} f(\wres) \frac{|\sigma|^3 Z_0^2}{\Im \sigma} \frac{1}{k_{\rm res}^2 d^2}.
    \label{eq:hrad_bnd}
\end{align}
Note that this is not a strict bound, but rather an indication of what is \emph{possible}, if the single-frequency bounds derived in the text can be reached over a typical plasmonic bandwidth (which is significantly narrower than the RHT flux rates seen in Fig.~4 of the main text). 

\Figref{figS2} shows the heat-transfer coefficient in graphene if \eqref{hrad_bnd} can be met. We fix the Fermi level at $\SI{0.6}{eV}$, consider two temperatures: $T = \SI{300}{K}$ and $T=\SI{1500}{K}$, for a resonant wavelength $\lres$ swept from $\SI{3}{\mu m}$ to $\SI{5}{\mu m}$. For the sake of comparison, we include the \emph{conductive} heat-transfer coefficient through air, taking the thermal conductivity to be $\kappa_{\rm air} = \SI{0.026}{W/m^2 \cdot K}$ (\citeasnoun{Haynes2013}). An exciting feature of \figref{figS2} is the length scale at which heat transfer may become dominated by radiative rather than conductive heat transfer. For $\SI{300}{K}$, this transition can occur at separation distances larger than $\SI{300}{nm}$, and for $\SI{1500}{K}$, the transition can happen beyond $\SI{800}{nm}$, separations orders of magnitude larger than those required with conventional designs.

\section{Graphene material figure of merit: second-order approximation}
A surprise in the material figure of merit of graphene is the extent to which interband contributions play a significant role in the peak magnitude of the response even at energies smaller than the Fermi level. The simplified expressions for graphene's material FOM given in Eq.~(9) of the main text are asymptotic expressions, and the low-energy expression is only valid for $\omega \ll \gamma$, where $\gamma$ is the small material loss rate. In this section, we derive a higher-order correction that more accurately describes a broader frequency range. For $\hbar\omega < 2\EF$, the low-temperature ($T \ll \EF / k_B$) approximations of the intra- and interband conductivities are
\begin{subequations}
\begin{align}
    \sigma_{\rm intra} &= \frac{i e^2}{4\pi\hbar} \frac{4\EF}{\hbar (\omega + i\gamma)} \\
    \sigma_{\rm inter} &= -\frac{i e^2}{4\pi\hbar} \ln\left( \frac{2\EF + \hbar(\omega + i\gamma)}{2\EF - \hbar\left(\omega + i\gamma\right)} \right).
\end{align}
\end{subequations}
A Taylor expansion in frequency (with small parameter $\hbar (\omega + i\gamma) / 2 \EF$) yields an inverse total conductivity of
\begin{align}
    \left(Z_0 \sigma\right)^{-1} \simeq -\frac{i}{\alpha} \frac{\hbar (\omega + i\gamma)}{4\EF} \left( 1 + \frac{\hbar^2 (\omega^2 - \gamma^2 + 2i\gamma\omega)}{4\EF^2}\right).
    \label{eq:inv_cond}
\end{align}
Inserting the inverse conductivity of \eqref{inv_cond} into the cross-section bound, Eq.~(6) of the main text, yields the approximate graphene bound:
\begin{align}
    \left(\frac{\sigma_{\rm ext}}{A}\right)_{\rm bound} &= \left[ \Re \left(Z_0 \sigma\right)^{-1} \right]^{-1} \nonumber \\
    &\simeq 4\alpha \left(\frac{E_{\rm F}}{\hbar \gamma}\right) - \alpha \frac{\hbar\gamma}{E_{\rm F}} \left[ 3 \left(\frac{\omega}{\gamma}\right)^2 - 1\right]
    \label{eq:secOrder}
\end{align}
\Eqref{secOrder} predicts a quadratic reduction in graphene's material figure of merit (and thus its response bounds) as a function of energy. As shown in \figref{figS2}, the quadratic dependence is a good approximation of the full local-response material conductivity for energies well below twice the Fermi level. Note that the frequency-dependent second term in \eqref{secOrder} arises entirely from \emph{inter}band contributions to the conductivity, which are a crucial limiting factors even at frequencies well below the Fermi level.
\begin{figure}
    \centering\includegraphics[width=0.48\textwidth]{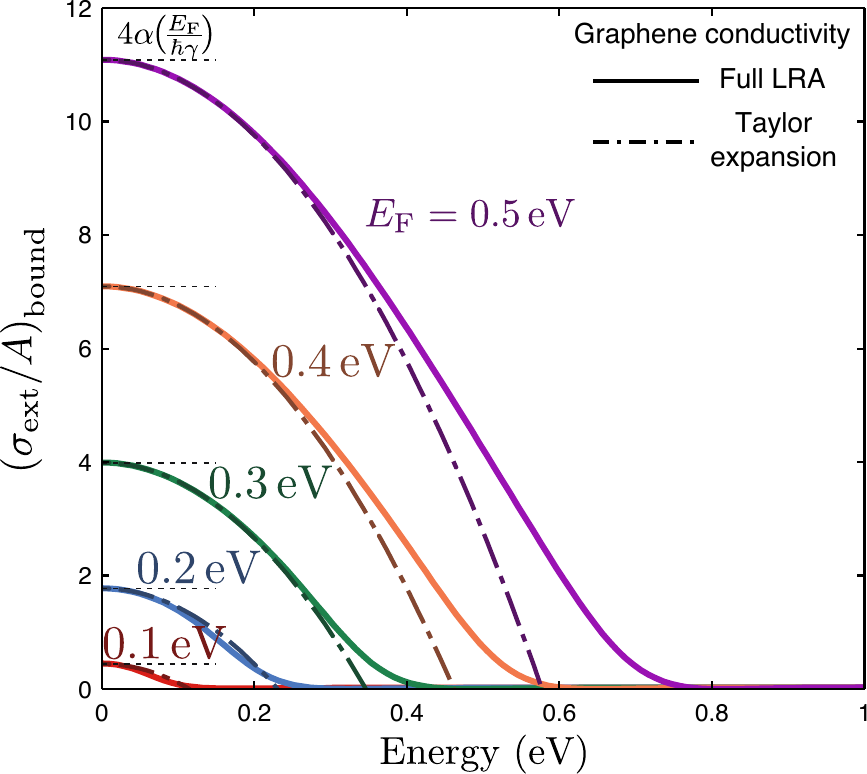}
    \caption{Comparison of the extinction bounds, $(\sigma_{\rm ext}/A)_{\rm bound}$, for graphene with the full local-response-approximation (LRA) conductivity (solid) and with the second-order approximation in \eqref{secOrder} (dash-dot). Even at frequencies below the Fermi level, inclusion of the interband terms, resulting in the quadratic dependence evident here, yields much better agreement than the intraband-only expression (dashed).}
    \label{fig:figS3}
\end{figure}

\section{Variational-calculus derivation of upper bounds}
Here we provide the intermediate mathematical steps in the derivation of the bounds that appear in Eqs.~(4--8) of the main text. For generality, we also accommodate the possibility of magnetic surface currents in addition to electric surface currents. We denote the fields as components of a six-vector $\psi$, 
\begin{align}
    \psi &= \begin{pmatrix}
        \Ev \\
        \Hv
    \end{pmatrix}
\end{align}
and the electric ($\Kv$) and magnetic ($\vect{N}$) surface currents as components of a six-vector $\phi$:
\begin{align}
    \phi &= \begin{pmatrix}
        \Kv \\
        \vect{N}
    \end{pmatrix}
\end{align}
Then we can write the absorption and extinction as the inner products of the fields and currents:
\begin{align}
    \Pabs &= \frac{1}{2} \Re \ip{\psi}{\phi} \\
    \Pext &= \frac{1}{2} \Re \ip{\psi_{\rm inc}}{\phi}
\end{align}
where the inner product is defined by $\ip{a}{b} = \int_A a^\dagger b \,{\rm d}A$. For the most general bounds in the main text, Eq.~(4), we assume only that the fields are currents are related by some linear operator $\Lke$,
\begin{align}
    \Lke \phi = \psi,
    \label{eq:const_eqn}
\end{align}
where we have generalized the $\Lke$ operator from the main text, to include magnetic currents.

The simplest bound to derive is the one for scattered power. We substitute the constitutive equation, \eqref{const_eqn}, in the equations for absorption and extinction, and write the scattered power as the difference between extinction and absorption:
\begin{align}
    \Pscat = \frac{1}{2} \left[\frac{1}{2} \ip{\psi_{\rm inc}}{\phi} + \frac{1}{2} \ip{\phi}{\psi_{\rm inc}} - \ip{\phi}{\left(\Re \Lke\right) \phi} \right]
    \label{eq:pscat_gen}
\end{align}
Note that by passivity $\Re \Lke$ is positive-definite (for a scalar isotropic conductivity, $\Re \Lke > 0$ is equivalent to $\Re \sigma > 0$). Thus the negative term in \eqref{pscat_gen} is a positive-definite quadratic function of the currents $\phi$, whereas the first two positive terms are only \emph{linear} in $\phi$. Thus $\Pscat$ is inherently bounded by constraints imposed by the optical-theorem form of the extinction. We can find the extremum by setting the variational derivative $\delta / \delta \phi^*$ equal to zero:
\begin{align}
    \vd{\Pscat} = \frac{1}{4} \psi_{\rm inc} - \frac{1}{2} \left(\Re \Lke\right) \phi = 0
\end{align}
which implies that the optimal currents are given by
\begin{align}
    \phi = \frac{1}{2} \left(\Re \Lke\right)^{-1} \psi_{\rm inc}
    \label{eq:phi_opt_s}
\end{align}
For these optimal currents, the scattered power is given by direct substitution of \eqref{phi_opt_s} into \eqref{pscat_gen}, yielding
\begin{align}
    \Pscat \leq \frac{1}{8} \ip{\psi_{\rm inc}}{\left(\Re \Lke\right)^{-1} \psi_{\rm inc}}.
    \label{eq:pscat_bnd}
\end{align}
\Eqref{pscat_bnd} is the magnetic-current generalization of the scattered-power component of Eq.~(4) in the main text. By similar variational derivatives, with a Lagrangian-multiplier approach to the constraint $\Pabs < \Pext$, the bounds on $\Pabs$ and $\Pext$ follow:
\begin{align}
    P_{\rm abs,ext} \leq \frac{1}{2} \ip{\psi_{\rm inc}}{\left(\Re \Lke\right)^{-1} \psi_{\rm inc}},
    \label{eq:pext_bnd}
\end{align}
with the only difference from the scattered-power bound being the extra factor of four (the $\beta$ term in the main text), which arises because maximization of absorption or extinction can fully ``saturate'' the constraint, i.e. $\Pabs = \Pext$. Similar saturation would yield no scattered power, and thus the scattered-power optimum occurs for $\Pabs = \Pscat = \frac{1}{2} \Pext$, at half the current level and thus one-fourth of the power level.

The next equation from the main text that we want to show the key steps for is Eq.~(7), the bound for the LDOS. In this case, we can consider a spatially local conductivity for the $\Lke$ operator, i.e., $\Lke = \sigmat^{-1}$. We henceforth do not consider magnetic currents, though the generalization is straightfoward. The bound for the LDOS takes exactly the same form as \eqreftwo{pscat_bnd}{pext_bnd}, for absorption, scattering, and extinction, but with a different prefactor to account for the free-space LDOS, $\rho_0$:
\begin{align}
    \frac{\rho_\alpha}{\rho_0} &\leq \beta_\alpha \frac{1}{\epso \omega} \frac{2\pi}{k^3} \sum_j \ip{\Ev_{{\rm inc},j}}{\left(\Re \sigmat^{-1}\right)^{-1} \Ev_{{\rm inc},j}} \nonumber \\
    &= \beta_\alpha \frac{1}{\epso \omega} \frac{2\pi}{k^3} \left\|\sigmat^\dagger \left(\Re \sigmat\right)^{-1} \sigmat\right\|_2 \sum_j \int_A  \left|\Ev_{{\rm inc},j}\right|^2 \,{\rm d}A
    \label{eq:rho_eqn}
\end{align}
where $j$ denotes the (random) orientation of the dipolar emitter, $\alpha$ denotes either the total, radiative, or nonradiatve LDOS, and $\beta_\alpha$ is $1$ for the total or nonradiative LDOS and $1/4$ for the radiative LDOS (and we have dropped an additive +1 factor for the radiative LDOS that is negligible in the near field). The surface $A$ of the 2D material can take any shape; because the integrand in \eqref{rho_eqn} is positive, we can find the planar surface passing through the point on $A$ that is closest to the emitter. Denoting this half space $\Gamma$, we know that
\begin{align}
    \int_A |\Ev_{\rm inc}|^2 \,{\rm d}A \leq \int_\Gamma |\Ev_{\rm inc}|^2 \,{\rm d}A
\end{align}
where the latter expression can be analytically evaluated due to its symmetry. [As discussed in the main text, other bounding surfaces (such as the closest spherical shell) can be used, instead of a half space, with the resulting difference only being different numerical prefactors. To determine the integral, we can use the fact that the sum of the squared electric field over all source-dipole orientations is given by the Frobenius norm of the dyadic electric-field Green's function:
\begin{align}
    \sum_j \left| \Ev_{{\rm inc},j}\right|^2 = \left\|\tens{G}_0\right\|_F^2 = \frac{k^6}{8\pi^2} \left[\frac{3}{\left(kr\right)^6} + \frac{1}{\left(kr\right)^4} + \frac{1}{\left(kr\right)^2} \right]
    \label{eq:GF0_sq}
\end{align}
which has contributions from $1/r^6$, $1/r^4$, and $1/r^2$ terms. The $1/r^2$ term represents a far field radiative contribution, which is dominated in the near field by higher-order terms. The integrals of $1/r^6$ and $1/r^4$ over the plane $\Gamma$ are
\begin{subequations}
\begin{align}
    \int_\Gamma \frac{1}{r^6}\,{\rm d}A &= \frac{\pi}{2d^4} \\
    \int_\Gamma \frac{1}{r^4}\,{\rm d}A &= \frac{\pi}{d^2}
\end{align}
\end{subequations}
where $d$ is the separation of the emitter from the plane $\Gamma$. Thus the integral over the Frobenius norm of the Green's function, excluding the far-field term, is
\begin{align}
    \int_\Gamma \left\|\tens{G}_0\right\|_F^2\,{\rm d}A = \frac{k^4}{8\pi} \left[ \frac{3}{2(kd)^4} + \frac{1}{(kd)^2} \right]
\end{align}
Inserting this value into the bound of \eqref{rho_eqn} yields:
\begin{align}
    \frac{\rho_\alpha}{\rho_0} \leq \beta_\alpha \left\|\sigmat^\dagger \left(\Re \sigmat\right)^{-1} \sigmat\right\|_2 \left[ \frac{3}{8(kd)^4} + \frac{1}{4(kd)^2}\right],
\end{align}
which is the LDOS bound of Eq.(7) in the main text, including the second-order term.

The final expression whose mathematical form we want to derive is the RHT bound of Eq.~(8) in the main text. As explained in the main text, and derived in \citeasnoun{Miller2015}, a bound on RHT can be developed by consideration of two scattering problems connected through (generalized) reciprocity. For two surfaces with conductivities $\sigmat_1$ and $\sigmat_2$, the bound is of the form
\begin{equation}
\begin{aligned}
    \Phi(\omega) \leq &\frac{2}{\pi \epso^2 \omega^2} \left\| \left(\Re \sigmat_1^{-1}\right)^{-1} \right\|_2 \left\| \left(\Re \sigmat_2^{-1}\right)^{-1} \right\|_2 \\
    &\times \int_{A_1} \int_{A_2} \left\| \tens{G}_0(\xv_1,\xv_2) \right\|_F^2 \, {\rm d}^2\xv_1 {\rm d}^2 \xv_2.
    \label{eq:PhiBnd}
\end{aligned}
\end{equation}
To complete the integral over the two 2D surfaces, we use the same ``bounding plane'' approach as for the LDOS. Now we need a double integral over $\Gamma_1$ and $\Gamma_2$, where $\Gamma_1$ and $\Gamma_2$ are the bounding planes for $A_1$ and $A_2$:
\begin{align}
    \int_{\Gamma_1} \int_{\Gamma_2} \frac{1}{r^6} = A \int_{\Gamma_2} \frac{1}{r^6} = \frac{\pi A}{2 d^4} \\
    \int_{\Gamma_1} \int_{\Gamma_2} \frac{1}{r^4} = A \int_{\Gamma_2} \frac{1}{r^4} = \frac{\pi A}{d^2}
\end{align}
where $A$ is the (infinite) area of the $\Gamma_1$ and $\Gamma_2$ surfaces, which could be pulled out of the integrals by symmetry. Inserting the integrals into the RHT bound expression in \eqref{PhiBnd} yields:
\begin{equation}
\begin{aligned}
    \Phi(\omega) \leq &\frac{k^2A}{4\pi^2} Z_0^2 \left\| \left(\Re \sigmat_1^{-1}\right)^{-1} \right\|_2 \left\| \left(\Re \sigmat_2^{-1}\right)^{-1} \right\|_2 \\
    & \times \left[ \frac{3}{2(kd)^4} + \frac{1}{4(kd)^2} \right].
\end{aligned}
\end{equation}
Recognizing that $k^2 A / 4\pi^2$ is precisely the blackbody flux rate~\cite{Joulain2005}, $\Phi_{\rm BB}$, we can write
\begin{equation}
\begin{aligned}
    \frac{\Phi(\omega)}{\Phi_{\rm BB}(\omega)} \leq & Z_0^2 \left\| \left(\Re \sigmat_1^{-1}\right)^{-1} \right\|_2 \left\| \left(\Re \sigmat_2^{-1}\right)^{-1} \right\|_2 \\
    & \times \left[ \frac{3}{2(kd)^4} + \frac{1}{4(kd)^2} \right],
\end{aligned}
\end{equation}
which is precisely the RHT bound of Eq.~(8) in the main text, except that here we allow for two different materials in the interaction, and we include the second-order distance term, proportional to $1/d^2$.

\section{Bounds in the presence of hydrodynamic nonlocality}
In the main text, we showed that in a general Maxwell-equation framework, hydrodynamic nonlocality \emph{cannot} increase maximum optical response, as any such nonlocal response is subject to the local-response bound. Here we show that under the additional assumption of \emph{quasistatic} response, which will almost always apply at the length scales for which nonlocal effects are non-negligible, the nonlocality necessarily \emph{reduces} the maximum achievable optical response in a given 2D material. In accord with typical hydrodynamic models~\cite{Mortensen2014}, we will work only with electric surface currents $\Kv$, driven by electric fields $\Ev$, related by Eq.~(10) of the main text, repeated here in compact notation:
\begin{align}
    -A \nabla\nabla \cdot \Kv + B\Kv = \Ev,
    \label{eq:HydroEqnSM}
\end{align}
where
\begin{subequations}
\begin{align}
    A &= \frac{i}{\epso \omega \omegap^2} \left(\beta^2 + D\left(\gamma - i\omega \right)\right), \\
    B &= \sigma_{\rm loc}^{-1},
\end{align}
\end{subequations}
$\sigma_{\rm loc}$ is the local surface conductivity, and $\beta^2 = (3/5)v_F^2$ (for Fermi velocity $v_F$) for both parabolic 2D materials as well as graphene. Note that one can define the plasma frequency $\omega_p$ using $\hbar k_{\rm F} / v_{\rm F}$ as the effective mass, yielding $\omega_p^2 = e^2 E_{\rm F} / (\pi \epso \hbar^2)$. In the presence of a hydrodynamic nonlocality, it is straightforward to write the absorbed power in terms of the currents $\Kv$:
\begin{align}
    P_{\rm abs} &= \frac{1}{2} \Re \int_A \cc{\Kv} \cdot \Ev \nonumber \\
                &= \frac{1}{2} \Re \int_A -A \cc{\Kv} \cdot \nabla \nabla \cdot \Kv + B \cc{\Kv} \cdot \Kv \nonumber \\
                &= \frac{1}{2} \int_A a \left( \nabla \cdot \cc{\Kv} \right) \left( \nabla \cdot \Kv\right) + b \cc{\Kv} \cdot \Kv,
    \label{eq:PabsNL}
\end{align}
where the second line follows from integration by parts and the no-spillout condition ($\hat{\vect{n}}\cdot \Kv = 0$), and $a$ and $b$ are the real parts of $A$ and $B$, respectively,
\begin{subequations}
\begin{align}
    a &= \Re(A) = \frac{D}{\epso \omegap^2}, \\
    b &= \Re(B) = \Re\left(\sigma_{\rm loc}^{-1}\right),
\end{align}
\end{subequations}
which are positive by the sign convention chosen in \eqref{HydroEqnSM}. The key insight to take away from \eqref{PabsNL} is that it is quadratic in $\Kv$ \emph{and} $\div \Kv$. Thus for nonlocal models, restrictions on the \emph{divergence} of the currents represent an additional constraint on maximal optical response. To have a non-trivial restriction on $\div \Kv$, there should also be a term in the extinction that is linear in $\div \Kv$. This is where the quasistatic approximation is useful. Quasisatic electromagnetism dictates that the incident field is the (negative) gradient of some potential $\phi_{\rm inc}$: $\Ev_{\rm inc} = -\nabla \phi_{\rm inc}$. Then, using integration by parts and the no-spillout condition once more, we can write the extinction in either of two equivalent ways:
\begin{align}
    \Pext^{(1)} &= \frac{1}{2} \Re \int_A \cc{\Ev_{\rm inc}} \cdot \Kv \label{eq:Pext1},\\
    \Pext^{(2)} &= \frac{1}{2} \Re \int_A \cc{\phi_{\rm inc}} \div \Kv \label{eq:Pext2}.
\end{align}
The first equation, \eqref{Pext1}, offers a constraint on the magnitude of $\Kv$, while the second equation, \eqref{Pext2}, offers a constraint on the magnitude of $\div \Kv$. Thus if we wish to maximize extinction, for example, it is subject to two constraints, $\Pabs < \Pext^{(1)}$ and $\Pabs < \Pext^{(2)}$, and we should maximize the \emph{minimum} of $\Pext^{(1)}$ and $\Pext^{(2)}$ (which are not necessarily equivalent since we do not impose the additional nonconvex constraint of satisfying quasistatic electromagnetism). Thus the maximal-extinction problem can be written as a ``maximin'' (negative of a minimax) convex problem
\begin{equation}
    \begin{aligned}
        & \max_{\Kv,\div\Kv} \min_{i \in \{1,2\}} & & \Pext^{(i)} \\
        & \text{such that}  & & \Pabs \leq \Pext^{(i)}.
    \end{aligned}
    \label{eq:MaxExt}
\end{equation}
Although \eqref{MaxExt} is nonsmooth (because of the absolute value), a standard transformation~\cite{Nocedal2006} yields an equivalent smooth optimization problem
\begin{equation}
    \begin{aligned}
        & \max_{\Kv,\div\Kv,x}  & & x \\
        & \text{such that}  & & x \leq \Pext^{(i)} \\
        & & & \Pabs \leq \Pext^{(i)},
    \end{aligned}
    \label{eq:MaxExtSmooth}
\end{equation}
where $i \in \{1,2\}$ and the constraints are all convex. At the extremum $\Pext^{(1)} = \Pext^{(2)}$, and standard optimization techniques (e.g., Lagrange multipliers) yield this optimal value:
\begin{align}
    \Pext \leq \frac{1}{2} \left[ \frac{\Re(\sigma_{\rm loc}^{-1})}{\int_A \left|\vect{\Ev_{\rm inc}}\right|^2} + \frac{D/(\epso \omegap^2)}{\int_A |\phi_{\rm inc}|^2} \right]^{-1}.
    \label{eq:fbndNL}
\end{align}
The bound on the right-hand side of \eqref{fbndNL} is a rate competition between the local-conductivity bound in the first term and a diffusion-constant-based bound in the second term that only arises from the hydrodynamic nonlocality. We can simplify the bound in the case of a plane wave.

Within the quasistatic approximation, an incident plane wave is represented by a constant vector field across/over the surface of the 2D material; for a polarization along $\hat{\mathbf{z}}$, i.e. for $\mathbf{E}_{\mathrm{inc}} = E_0\hat{\mathbf{z}}$, the associated potential is $\phi_{\mathrm{inc}} = -E_0z$. If the ``radius'' of the scatterer (more precisely, its smallest bound sphere in the polarization direction) is given by $r$, we can simplify the integral of $|\phi_{\rm inc}|^2$ via the inequality
\begin{align}
    \int_A |\phi_{\rm inc}|^2 = |E_0|^2 \int_A z^2 = |E_0|^2 \left\langle z^2 \right\rangle A \leq |E_0|^2 r^2 A,
\end{align}
where $\langle \cdot \rangle$ denotes an average over the area of the scatterer. In terms of the cross-section, $\sigma_{\rm ext} = \Pext / (|E_0|^2 / 2Z_0)$, the expression of \eqref{fbndNL} is bounded above by
\begin{align}
    \frac{\sigma_{\rm ext}}{A} \leq \left[ \left(Z_0 \frac{\left| \sigma_{\rm loc} \right|^2}{\Re \sigma_{\rm loc}}\right)^{-1} +  \left(\frac{r^2}{\ell_D^2}\right)^{-1} \right]^{-1},
    \label{eq:extBndNL}
\end{align}
where $\ell_D = \sqrt{\frac{cD}{\omegap^2}}$ is a normalized diffusivity that we can interpret as a plasmonic ``diffusion'' length. \Eqref{extBndNL} has an appealing, intuitive interpretation: the cross-section of a scatterer is bounded above by a combination of the local-conductivity bound and a nonlocal contribution proportional to the square of the ratio of the size of the scatterer to the ``diffusion'' length. Thus as the size of the particle approaches $\ell_D$, and goes below it, there is a significant reduction in the maximal optical response.

Because the local density of states (LDOS) is proportional to \eqref{Pext1}, but with the replacement $\cc{\Ev_{\rm inc}} \rightarrow \Ev_{\rm inc}$ (\citeasnoun{Miller2016}), the equivalent LDOS bound is exactly \eqref{fbndNL}, with additional numerical prefactors and the caveat that $\Ev_{\rm inc}$ is now rapidly decaying in space. The $1/r^3$ decay of the incident field is responsible for the $1/d^4$ distance dependence of the local-conductivity LDOS bound, Eq.(7), in the main text. But the incident-field potential, $\phi_{\rm inc}$, decays less rapidly, with scaling $\sim 1/r^2$. Thus $\int_A |\phi_{\rm inc}|^2 \sim 1/d^2$, a dramatic reduction from the $1/d^4$ scaling for a local conductivity. The crossover from the $1/d^4$ term being dominant in the bound to the $1/d^2$ term being dominant occurs when the separation distance $d$ is of the same order of magnitude as the diffusion length $\ell_D$. Exploration of the $1/d^2$ scaling in various relevant materials and geometries would be interesting future work.

\section{LDOS above a planar conducting sheet}
In this section we analytically derive the LDOS above a planar conducting sheet. We show that the envelope of the peak LDOS has $1/d^3$ scaling when dominated by a single resonance, whereas it has a $1/d^4$ scaling, and comes within a factor of two of the LDOS bounds of Eq.~(7) in the main text, when it arises from a ``lossy-background'' contribution. The LDOS above any structure with translational and rotational symmetry is given by
\begin{align}
    \rho(\omega) = \int \rho(\omega,k_p) \, {\rm d}k_p
\end{align}
where $k_p$ is the magnitude of the surface-parallel component of the wavevector. In the near field ($k_p \gg k_0$), for $p$-polarized waves (e.g., surface plasmons), $\rho(\omega,k_p)$ is given by
\begin{align}
    \rho(\omega, k_p) = \frac{k_0}{2\pi^2 c} \frac{k_p^2}{k_0^2} e^{-2k_p z} \Im r_p
    \label{eq:rhoInt}
\end{align}
where $r_p$ is the $p$-polarized (TM) reflection coefficient. For a 2D material with surface conductivity $\sigma$, $r_p$ is given by
\begin{align}
    r_p &\approx \frac{i \sigma k_p}{2\varepsilon_0 \omega + i\sigma k_p} \label{eq:rp_sigma} \\
    &= \frac{k_p}{k_p - \xi}
\end{align}
where $\xi = i 2\epso \omega / \sigma$. Thus the imaginary part of the reflection coefficient is
\begin{align}
    \Im r_p = \frac{k_p \xi''}{(k_p - \xi)' + (\xi'')^2},
    \label{eq:ImR}
\end{align}
where the single and double apostrophes indicate real and imaginary parts, respectively. The variable $\xi(\omega)$ encodes the material conductivity. For single-resonance-dominant response, the wavevector integral of \eqref{rhoInt} will be dominated by a narrow peak in the imaginary part of the reflection coefficient, i.e. \eqref{ImR}, where $k_p \approx \xi'$. Conversely, for a highly lossy background, for which $\Re \sigma \gg |\Im \sigma|$ and thus $\Im \xi \gg |\Re \xi|$, the contribution of $\Im r_p$ to the integrand in \eqref{PhiBnd} will be roughly constant. We treat the two cases separately.
\subsection{Pole contribution to the LDOS}
    As discussed above, the imaginary part of the reflection coefficient will be sharply peaked around $k_p \approx \Re \xi(\omega)$ in the case of a single resonance dominating the response. Then the peak value of $\Im r_p$, as a function of wavevector, will be
    \begin{align}
        \max \Im r_p \approx \frac{k_p}{\xi''}
    \end{align}
and the width of the peak will be $\Delta k_p \approx 2 \xi''$. If we denote $k_{p0}$ as the peak wavevector at which $\Im r_p$ takes its maximum value, and assume a Lorentzian lineshape for $\Im r_p$, then we can approximate the $k_p$-dependent terms in the integral of \eqref{rhoInt} by
\begin{align}
    \int k_p^2 e^{-2k_p z} \Im (r_p) \, {\rm d}k_p &\approx k_{p0}^2 e^{-2k_{p0} z} \int \Im (r_p) \, {\rm d}k_p \nonumber \\
    &\approx k_{p0}^2 e^{-2k_{p0} z} \frac{\pi}{2} \Im \left[ r_p(k_{p0}) \right] \Delta k_p \nonumber \\
    &= \pi k_{p0}^3 e^{-2k_{p0} z}
\end{align}
Thus we can write the full LDOS, $\rho(\omega)$, as
\begin{align}
    \rho(\omega) = \rho_0(\omega) \frac{k_{p0}^3}{k_0^3} e^{-2 k_{p0} z},
\end{align}
where $\rho_0(\omega)$ is the electric-only free-space LDOS, $\rho_0 = k_0^2 / 2\pi^2 c$. We note that the optimal frequency, and thus the optimal $k_{p0}$, changes as a function of $z$, with the optimal $k_{p0}$ given by $k_{p0} = 3/2z$. Replacing the height $z$ with the separation distance $d$, we can write
\begin{align}
    \max_\omega \frac{\rho(\omega)}{\rho_0(\omega)} &\approx \pi \left(\frac{3}{2e}\right)^3 \frac{1}{(k_0d)^3} \nonumber \\
                                                    &\approx \frac{1}{2(k_0 d)^3}.
                                                    \label{eq:ldos_pole}
\end{align}
The expression given by \eqref{ldos_pole} quantitatively predicts the short-distance and low-frequency behavior of the LDOS in Fig.~3 of the main text.

\subsection{Lossy-background contribution to the LDOS}
The lossy-background contribution to the LDOS exhibits a different mathematical structure. Instead of $\Im r_p$ being sharply peak around a single resonance, $\Im \xi \gg |\Re \xi|$, and the imaginary part of the reflectivity is nearly constant over wavevector:
\begin{align}
    \Im r_p \approx \frac{k_p}{\xi''}
\end{align}
for all $k_p$ (that are not so large as to be inaccessible at a finite separation distance). Thus $\Im r_p$ can be taken out of the integral for $\rho$, \eqref{rhoInt}, which is then given by 
\begin{align}
    \int k_p^2 e^{-2 k_p z} \Im r_p \, {\rm d}k_p &\approx \frac{1}{\xi''} \int k_p^3 e^{-2 k_p z} \, {\rm d}k_p \nonumber \\
                                                &\approx \frac{1}{\xi''} \frac{3}{8z^4},
\end{align}
where we have kept ony the lowest-order term in $1/z$. Writing out $\xi'' = 2\epso \omega / \Re \sigma$, straightforward algebra yields:
\begin{align}
    \frac{\rho(\omega)}{\rho_0(\omega)} \approx \frac{3}{16} \left(Z_0 \Re \sigma\right) \frac{1}{(k_0 d)^4}
    \label{eq:rhoLB}
\end{align}
for emitter--material separation distance $d$. We see that in the limit $\Re \sigma \gg |\Im \sigma|$, which is a prerequisite for the lossy-background contribution to dominate, \eqref{rhoLB} is exactly a factor of 2 smaller than the general LDOS bound that appears in Eq.~(7) of the main text. The factor of 2 stems from the factor of 2 in the denominator of \eqref{rp_sigma}, which itself arises from the equal interactions of a 2D material with the exterior regions on either side of its surface. \Eqref{rhoLB} quantitatively predicts the LDOS in the moderate-separation and large-energy regimes of Fig.~3 of the main text.

\bibliography{/Users/odm5/Library/texmf/bibtex/bib/library,/Users/odm5/Library/texmf/bibtex/bib/journalshort,/Users/odm5/Library/texmf/bibtex/bib/nonotes,/Users/odm5/Library/texmf/bibtex/bib/My_Pubs_abbr}